\documentclass[sigconf, nonacm=true]{acmart}

\usepackage{booktabs}
\usepackage{paralist}
\usepackage[caption=false]{subfig}

\usepackage[group-separator={,}, group-four-digits, per-mode=symbol, binary-units]{siunitx}

\usepackage[english]{babel}
\usepackage{microtype}

\usepackage{color}
\usepackage{xspace}
\usepackage{eurosym}

\usepackage{graphicx}
\usepackage[ruled,vlined]{algorithm2e}
\usepackage{threeparttable, tablefootnote}

\newcommand{\ie}{i.\@\,e.\@\xspace}
\newcommand{\eg}{e.\@\,g.\@\xspace}
\newcommand{\cf}{cf.\@\xspace}
\newcommand{\etal}{et~al.\@\xspace}

\newcommand{\Paragraph}[1]{\smallskip\noindent{\bf #1.}}

\definecolor{Gray}{gray}{0.5}
\PassOptionsToPackage{hyphens}{url}\usepackage{hyperref}

\AtBeginDocument{%
  \providecommand\BibTeX{{%
    \normalfont B\kern-0.5em{\scshape i\kern-0.25em b}\kern-0.8em\TeX}}}

\begin{document}

\title{Mirror Mirror on the Wall: Wireless Environment Reconfiguration Attacks Based on Fast Software-Controlled Surfaces}

\author{Paul Staat}
\affiliation{%
  \institution{Max Planck Institute for Security and Privacy}
  \city{Bochum}
  \country{Germany}
}
\author{Harald Elders-Boll}
\affiliation{%
  \institution{TH Köln – University of Applied Sciences}
  \city{Cologne}
  \country{Germany}
}
\author{Markus Heinrichs}
\affiliation{%
  \institution{TH Köln – University of Applied Sciences}
  \city{Cologne}
  \country{Germany}
}
\author{Christian Zenger}
\affiliation{%
  \institution{PHYSEC GmbH}
  \city{Bochum}
  \country{Germany}
}
\author{Christof Paar}
\affiliation{%
  \institution{Max Planck Institute for Security and Privacy}
  \city{Bochum}
  \country{Germany}
}

\begin{abstract}
The intelligent reflecting surface (IRS) is a promising new paradigm in wireless communications for meeting the growing connectivity demands in next-generation mobile networks. IRS, also known as software-controlled metasurfaces, consist of an array of adjustable radio wave reflectors, enabling smart radio environments, e.g., for enhancing the signal-to-noise ratio~(SNR) and spatial diversity of wireless channels. Research on IRS to date has been largely focused on constructive applications.

In this work, we demonstrate for the first time that the IRS provides a practical low-cost toolkit for attackers to easily perform complex signal manipulation attacks on the physical layer in real time. We introduce the environment reconfiguration attack~(ERA) as a novel class of jamming attacks in wireless radio networks. Here, an adversary leverages the IRS to rapidly vary the electromagnetic propagation environment to disturb legitimate receivers. The IRS gives the adversary a key advantage over traditional jamming: It no longer has to actively emit jamming signals, instead the IRS reflects existing legitimate signals. In addition, the adversary doesn't need any knowledge about the legitimate channel. 
We thoroughly investigate the ERA in wireless systems based on the widely employed orthogonal frequency division multiplexing~(OFDM) modulation. 
We present insights into the attack through analytical analysis, simulations, as well as experiments. Our results show that the ERA allows to severely degrade the available data rates even with reasonably small IRS sizes. Finally, we implement an attacker setup and demonstrate a practical ERA to slow down an entire \mbox{Wi-Fi} network.  
\end{abstract}

\maketitle

\section{Introduction}

Part of the ever-evolving digital landscape is growing demand for wireless connectivity at high data rates and low latency. 
In addressing this need, increasingly sophisticated mobile communication networks are being deployed. In particular, we are in the midst of the worldwide roll-out of 5G networks, which are the key-enablers for emerging applications such as, \eg, autonomous driving, smart cities, smart grids, and immersive entertainment \cite{andrewsWhatWill5G2014, guptaSurvey5GNetwork2015, agiwalNextGeneration5G23}. Such applications will lead to an increased dependency on a wireless infrastructure with high availability and high attack resistance. Specific to wireless networks is jamming of radio signals, which leads to denial of service and can pose a serious threat to, \eg, cellular networks such as 4G and 5G~\cite{lichtmanLTELTEAJamming2016, girkeResilient5GLessons2019, arjouneSmartJammingAttacks2020}.

\begin{figure}
\centering
\includegraphics[width=0.8\linewidth]{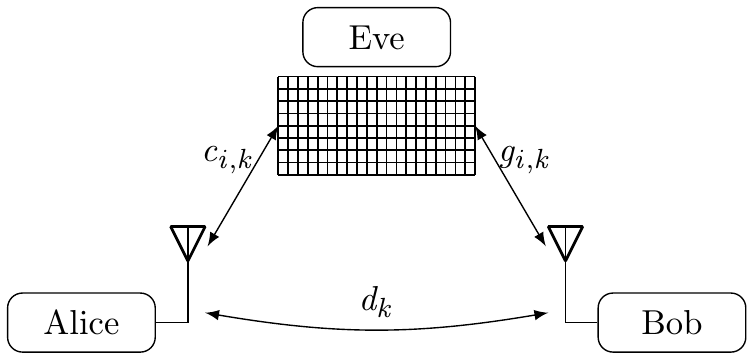}
\caption{Illustration of the ERA setting where the attacker Eve uses an IRS to gain partial control over the wireless channel between legitimate parties Alice and Bob. $c_{i,k}$ and $g_{i,k}$ are the channels to (and from) the IRS, $d_k$ is the direct (non-IRS) channel, with the $k^{th}$ OFDM subcarrier and $i^{th}$ IRS element.}
\label{fig:system_model}
\end{figure}

Next-generation wireless networks make use of sophisticated communication technologies such as massive MIMO (\sloppy massive multiple-input and multiple-output), which is now realized with 5G~\cite{bjornsonMassiveMIMOReality2019}. An even more recent example for a technological advance are \textit{intelligent reflecting surfaces}~(IRS)~\cite{wuSmartReconfigurableEnvironment2020}. IRS consist of an array of electronically adjustable reflectors with respect to radio waves. IRS enable \textit{smart radio environments}~\cite{liaskos_novel_2019, renzoSmartRadioEnvironments2019} to, \eg, enhance the wireless radio channel quality in terms of signal-to-noise ratio~(SNR)~\cite{kaina_shaping_2015} or spatial diversity~\cite{del_hougne_optimally_2019}. 

However, the IRS is also a novel attacker tool for malicious purposes --- an issue that has received only little attention as of yet. In this work, we show that time-varying IRS allow to disrupt wireless communications by (smart) reflecting radio signals originating from the legitimate parties. We introduce the environment reconfiguration attack~(ERA), which can be viewed as a novel class of practical, low-cost, and low-complexity jamming attacks. The essence of the ERA lies in high-speed IRS reconfigurations, which are digitally controlled by the attacker Eve. In effect, the wireless propagation environment, \ie, the wireless channel, between the communication parties Alice and Bob (cf.~Fig.~\ref{fig:system_model}) exhibits exceptionally fast and instantaneous changes that otherwise do not occur in nature. In turn, severe variations are applied to signals coming from the legitimate transmitter which disturb the intended receiver. A key difference to traditional jamming attacks is that the attacker does not actively emit a jamming signal but merely reflects signals generated by a victim party. Accordingly, the ERA leads to correlated interference and dramatically simplifies the implementation of such attacks~\cite{lichtmanCommunicationsJammingTaxonomy2016}, as the attacker neither needs an RF transmitter nor a receiver. Unlike previous work~\cite{lyuIRSBasedWirelessJamming2020}, the ERA does not require the attacker to have any channel knowledge and only rudimentary knowledge (such as the modulation scheme) about the communication system. This crucial relaxation allows us to demonstrate the first real-world jamming attack based on IRS.

In this paper, we show that the IRS is a practical and low-cost attacker tool, enabling the ERA. We  investigate the attack using orthogonal frequency division multiplexing~(OFDM) which is widely used in modern wireless networks, including 4G, 5G, and \mbox{Wi-Fi}. We perform a thorough theoretical analysis to explain the fundamental attack mechanisms. Furthermore, we show simulation results that allow us to  characterize the attack requirements on signal power, distances and IRS dimensions. Finally, we implement an attacker setup and demonstrate a practical ERA, slowing down an entire wireless network. Our results show that the attack works with reasonably small IRS sizes, notably the used IRS has dimensions \SI{40}{\cm}~$\times$~\SI{16}{\cm}. Moreover, we provide a practical IRS optimization algorithm to enhance the attack performance.

In summery, building upon the advent of  IRS, we introduce a new class of practical jamming attacks which are low-cost and can easily be deployed in many wireless scenarios. 
The paper at hand contains the following key contributions:
\begin{compactitem}

    \item We propose the environment reconfiguration attack~(ERA) as a novel class of jamming attacks, based on low-cost IRS.
    
    \item We present a theoretical analysis explaining how the ERA affects OFDM communications.
    
    \item We show comprehensive simulation results to determine the attacker requirements on signal power, distances and IRS dimensions.
    
    \item We demonstrate a practical ERA on commodity \mbox{Wi-Fi} using a low-cost IRS prototype, allowing to substantially reduce the wireless throughput in the entire network.
    
    \item We present an IRS optimization algorithm to further enhance the ERA jamming performance. 
\end{compactitem}

\section{Background}

In this section, we provide technical background on the IRS, jamming attacks, and OFDM communications. 

\subsection{Intelligent Reflecting Surface}

An IRS is a synthetic planar structure with digitally reconfigurable reflection properties of electromagnetic~(EM) waves. In wireless communications, the IRS is a rather new concept that has evolved from physics research on metamaterials and metasurfaces~\cite{kaina_shaping_2015} which are tailored to enable non-standard EM wave field manipulations. More recently, the evolutionary step from the metasurface to the IRS has been made: Metasurface designs have been drastically simplified and became digitally controllable. An IRS consists of many distributed identical unit cells, each of which reflects impinging EM waves. Most importantly, the complex reflection coefficient of each element across the surface is individually programmable, allowing to influence the wireless channel of communication parties (see Fig.~\ref{fig:system_model}). Practical IRS designs are often targeted to adjust only the signal phase with quantization as low as \SI{1}{bit}~\cite{yang1Bit10Times2016}. Thus, the IRS provides a simple digital interface towards the physical layer of wireless communications and enables what is coined \textit{smart radio environments}~\cite{liaskos_novel_2019} with novel applications such as, \eg, optimization of the signal-to-noise ratio~(SNR)~~\cite{basar_wireless_2019} or spatial diversity~\cite{del_hougne_optimally_2019}. Since only ambient signals are reflected, the IRS is inherently energy efficient and does not require active RF chains. Thus, IRS have low hardware complexity since manufacturing requires standard microstrip technology on low-cost printed circuit board~(PCB) substrate. Currently, the IRS is in discussion to complement future wireless infrastructure on a large scale in wireless networks beyond 5G~\cite{yang6GWirelessCommunications2019}.

\subsection{Jamming}
Wireless communication relies on a broadcast medium that must be shared between many users. In principle, each user is free to transmit at any time and thus, signals are by definition subject to interference. Instead of just the desired signal, a receiver then additionally picks up an unwanted signal, disrupting the intended communication. Despite regularly occurring interference from other user's communications, malicious parties can also launch \textit{jamming attacks}. Here, an attacker deliberately produces interference to disable the communication of targeted users. Jamming attacks can be classified into a variety of different categories, including the type of interference and the strategy to trigger emission of the interfering signal~\cite{groverJammingAntijammingTechniques2014}. A jammer may use noise signals, constant tones, or even valid waveforms. Attackers can apply constant jamming or act reactively in order to disable only selected parts of the victim communication, such as physical control channels~\cite{girkeResilient5GLessons2019}.

\subsection{Orthogonal frequency division multiplexing (OFDM)}
Due to its unique properties, OFDM has become one of the most important and widely used modulation techniques in wireless networks~\cite{chiuehBasebandReceiverDesign2012, goldsmith_wireless_2005}. Most importantly, OFDM can cope with multipath signal propagation easily. In order to push data rates, wide channel bandwidths need to be used. However, when transmitting a wide-bandwidth signal over a wireless link, it will most likely experience some form of frequency selective attenuation due to fading from multipath signal propagation. OFDM divides a wide bandwidth into numerous independent (say, orthogonal) narrowband channels, \ie, subcarriers, and can thus handle frequency selective channels at low computational complexity. Taking the concept to the next level, OFDM based multiple access~(OFDMA) schemes assign different subcarriers to different users. Finally, the modulation and demodulation of OFDM are elegantly handled using an efficient (inverse) fast Fourier transform~(FFT). Today, OFDM has become the definitive transmission scheme for broadcasting, \eg, DAB and DVB, cellular systems, \eg, 4G and 5G, and personal networks, \eg, \mbox{Wi-Fi}.

\section{Related Work}

In this section, we summarize the relevant  literature on IRS and jamming attacks, and also describe how our work differs from previous proposals.

\Paragraph{Intelligent reflecting surface}
The IRS has been widely recognized as a potential major innovation in wireless communications and has stimulated much research activity recently. Hence, there is a manifold literature now. Regarding key concepts and literature reviews, we refer to numerous overview works~\cite{basar_wireless_2019, wuSmartReconfigurableEnvironment2020, wuIntelligentReflectingSurface2021, liaskos_novel_2019}.

To the best of our knowledge, previous works on IRS in a security context focus on theoretical aspects. Most notably, Lyu~\etal~\cite{lyuIRSBasedWirelessJamming2020} proposed the IRS for minimizing the signal power received by a victim party for jamming. We further elaborate the similarities and differences to our work towards the end of this section. Several works, \eg, \cite{cuiSecureWirelessCommunication2019} and \cite{chenIntelligentReflectingSurface2019}, provide analytical and simulation results in the context of physical layer security assisted by an IRS. Huang and Wang~\cite{huangIntelligentReflectingSurface2020} discuss a pilot contamination attack using an IRS to increase signal leakage by reflecting pilot signals. In~\cite{yangIntelligentReflectingSurface2021}, the authors pursue IRS to be used as a mitigation for active jamming attacks.

In the following we give examples for studies including practical IRS demonstrations with a focus on improving wireless communication. An early work from 2014 is~\cite{kaina_shaping_2015}, where the authors demonstrate wave field shaping. Work from 2019~\cite{del_hougne_optimally_2019} has shown that IRS are capable of enhancing spatial diversity. Arun and Balakrishnan in 2020~\cite{arunRFocusBeamformingUsing2020} demonstrated a large prototype IRS with $3200$~elements for passive beamforming applications. In recent work of Pei~\etal~\cite{peiRISAidedWirelessCommunications2021}, an IRS is used to achieve substantial channel improvements, enabling a long-range communication field trial over \SI{500}{m}. Several works report practical IRS designs, \eg,~\cite{yang1Bit10Times2016, humReconfigurableReflectarraysArray2014, yangProgrammableMetasurfaceDynamic2016}.

\Paragraph{Jamming attacks}
The literature widely recognizes jamming attacks as a risk to the reliability of wireless communications. Several works have pointed out the threat of jamming against 4G~\cite{lichtmanLTELTEAJamming2016, girkeResilient5GLessons2019} and 5G~\cite{arjouneSmartJammingAttacks2020} networks. Grover~\etal~\cite{groverJammingAntijammingTechniques2014} provide an overview on different jamming strategies, localization and detection techniques, and countermeasures. However, the ERA does not fit any of the reported categories properly. Poisel gives a highly comprehensive overview on all classes of jamming in his book~\cite{poiselModernCommunicationsJamming2011}. Lichtman~\etal~\cite{lichtmanCommunicationsJammingTaxonomy2016} provide a taxonomy for jamming attacks by defining four attacker capabilities \textit{time correlation}, \textit{protocol awareness}, \textit{ability to learn}, and \textit{signal spoofing}. Following their categories, the ERA may be labeled as a partially time-correlated jammer. However, unlike the author's category-based conjecture, the ERA is a low-complexity attack. Hang~\etal~\cite{hangPerformanceDSSSRepeater2006} investigate repeater jamming against direct sequence spread spectrum (DSSS). The ERA may indeed be seen as a special case of repeater jamming, as a reflection of the signal in fact is a time-varying copy of the legitimate signal. Thus, the ERA is conceptually related. In the ERA, however, the attacker eliminates RF receiver and transmitter chains and processing delays. Pöpper~\etal~\cite{popperJammingResistantBroadcastCommunication2009} report a method to achieve jamming-resistant broadcast communications without shared keys. The authors comment on the repeater jammer which could circumvent their security assumptions in some cases and also point to processing delays. For our IRS-based approach, however, processing delays vanish. Clancy~\cite{clancyEfficientOFDMDenial2011} has pointed out that OFDM communications can be efficiently disrupted by jamming or nulling of pilot signals for channel estimation. The ERA now provides a simple method to realize the manipulation of the OFDM equalizer. Also, many works pursue detection of jamming, examples include~\cite{strasserDetectionReactiveJamming2010, chiangCrossLayerJammingDetection2011, lyaminRealTimeDetectionDenialofService2014}. A different body of work examines helpful aspects of jamming, \eg, to provide confidentiality~\cite{wenboshenAllyFriendlyJamming2013}. However, Tippenhauer~\etal~\cite{tippenhauerLimitationsFriendlyJamming2013} have shown that jamming for confidentiality has fundamental security limitations.

\Paragraph{Differentiation from previous work}
The general idea of maliciously using an IRS for jamming was first proposed by Lyu~\etal~\cite{lyuIRSBasedWirelessJamming2020} in 2020, albeit in a very different manner that we believe results in a much lower practicality than the ERA. 

The approach of \cite{lyuIRSBasedWirelessJamming2020} is based on an IRS to minimize the signal power received by a victim party -- a method opposite to the classical IRS-based SNR improvement. Here, the superposition of the direct signal and the malicious IRS signal shall 
result in \textit{destructive interference}, \ie, the IRS signal is to be a phase-exact cancellation signal. However, finding a specific IRS configuration to meet this goal is  non-trivial. Addressing this issue, the authors formulate an optimization scheme to obtain a corresponding IRS configuration from the channel states $c_{i,k}$, $g_{i,k}$, and $d_k$, \cf~Fig.~\ref{fig:system_model}. Thus in this approach the attacker needs to have full knowledge of all involved channel states. Unfortunately for an attacker, $d_k$ can only be found by the victim parties and obtaining $c_{i,k}$ and $g_{i,k}$ is infeasible (without a large number of additional RF receivers at the attacker's IRS), as recognized in the literature~\cite{basar_wireless_2019, wuSmartReconfigurableEnvironment2020, wuIntelligentReflectingSurface2021}. 

In contrast, the ERA approach presented in this paper works entirely different, thereby eliminating the unrealistic requirement of channel knowledge for the attacker. Crucially, the attack leverages the IRS to rapidly toggle between (two) effective wireless channels. In particular, we address OFDM receivers which get disturbed by the unnatural switching between channel states, \eg, partly due to adaptive behavior. Our goal is not the minimization of the signal reception of one or both of the ERA channels. Rather, the ERA exploits signal changes from the difference between the two ERA channels as a source of interference. Thus, the attack neither requires synchronization or phase-exact knowledge of all channels, and thereby avoids a location-dependent attack performance (signal phase changes by movement), as our experimental results show.

In order to compare the two attack strategies, we would like to point out that a cancellation approach~\cite{lyuIRSBasedWirelessJamming2020} is equivalent to reducing the SNR -- an aspect that we readily cover in our simulations in Section~\ref{sec:sim_jsr}, showing that the ERA can achieve substantially increased jamming performance.

\section{Attack Overview}

\begin{figure}
\centering
\includegraphics[width=1.0\linewidth]{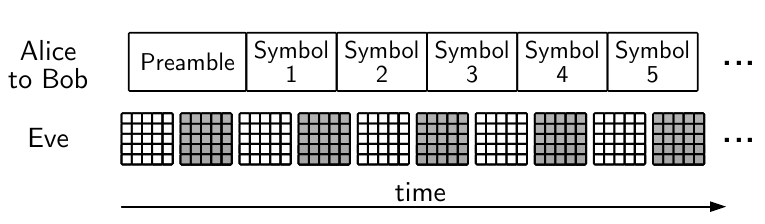}
\caption{Illustration of the ERA, indicating the legitimate communication and the adversarial IRS operation. The attacker toggles the IRS configuration rapidly to disturb the legitimate receiver.}
\label{fig:attack_timing}
\end{figure}

\Paragraph{Parties} 
In this work, we consider a physical layer attacker Eve trying to disrupt the wireless radio communication of two legitimate parties Alice and Bob who deploy a conventional OFDM-based wireless communication system. Thus, Alice and Bob may use \mbox{Wi-Fi}, 4G, or 5G and could represent a base-station and an end-user, respectively. 
The attacker Eve has full control over an IRS which is part of the wireless propagation channel between Alice and Bob. Eve is capable of applying custom configurations to the IRS at update rates comparably to the symbol rate used by Alice and Bob. Apart from that, we grant the attacker basic wireless eavesdropping capabilities, \ie, the attacker possesses a wireless receiver and can receive and demodulate signals of Alice and Bob. However, Eve does not have a wireless transmitter and thus cannot transmit any signals on itself.

Finally, our system and attacker model is illustrated in Fig.~\ref{fig:system_model}. Note that the attacker operates at the physical layer and therefore we do not need to take the cryptography applied at the upper layer of the user's communication into account.

\Paragraph{Attack and overview of investigation}
In the ERA, the attacker Eve uses a software-controlled surface, \ie, an IRS, to rapidly vary the wireless radio channel between Alice and Bob. This yields fast and instantaneous variations in the legitimate signals that normally would not occur in nature. Disturbed by the anomalous signal behavior, the intended receiver fails to correctly demodulate the incoming signals, leading to a denial of service. In this work, we design an ERA against OFDM communications by rapidly toggling between two distinct IRS configurations. An illustration of the corresponding attacker action is shown in Fig.~\ref{fig:attack_timing}. Compared to classical jamming attacks, the ERA allows attackers to silently disable the wireless communications of victim parties, \ie, the attacker does not actively generate a jamming signal. Instead, it manipulates signals transmitted by Alice and Bob during propagation.

We begin our investigations by examining the fundamental attack mechanisms in an analytical analysis~(Section~\ref{sec:analytical_analysis}). Here, we lay the foundations of the attack and show that ERA-induced fast channel variations are harmful for wireless OFDM communication. We then turn to a simulation model (Section~\ref{sec:sim_results}) of an end-to-end wireless OFDM link. From the simulation, we deduce several key factors of the attack, such as, \eg, signal power and attacker distances. For both theoretical analysis and simulations, we abstract the effect of the adversarial IRS as a time-varying signal component and omit the impact of specific IRS patterns. Finally, we use a practical IRS implementation to design and evaluate real-world ERAs to demonstrate successful jamming attacks~(Section~\ref{sec:experiments}). In the first and simplest variant, we rapidly toggle the IRS patterns by either setting all elements to '0' or '1'. This attack is of remarkably low complexity and requires nothing more than a certain proximity between the attacker and a victim party. The second attack variant is more advanced and includes an optional setup phase where the attacker optimizes the two IRS patterns to increase the jamming efficiency. This procedure incorporates the channel state information~(CSI) from Alice and Bob, as provided by CSI feedback signals in existing wireless standards.

\section{Theoretical Analysis}
\label{sec:analytical_analysis}

In this section, we present a theoretical analysis of the mechanisms underlying the ERA against OFDM communications. We outline that the ERA affects channel equalization from outdated channel estimations and subcarrier orthogonality.

\subsection{Modelling Preliminaries}
We begin our considerations by introducing the models for the legitimate OFDM communications and the IRS attacker.

\subsubsection{OFDM}
\begin{figure}
\includegraphics[width=1.0\linewidth]{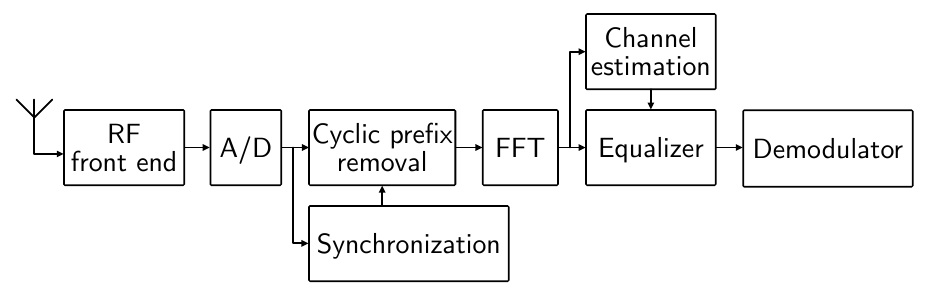}
\caption{Block-diagram of a typical OFDM receiver architecture.}
\label{fig:ofdm_receiver}
\end{figure}

We assume that Alice and Bob generate their RF transmit signals using a modulator fed by conventional complex-valued in-phase and quadrature~(IQ) baseband signals~\cite{goldsmith_wireless_2005}. The baseband signals for OFDM are generated by taking the inverse discrete Fourier transform of a block of $K$~complex modulated data symbols $X_k[n]$ for all $k= 0,\ldots, K-1$ subcarriers, yielding the $n^{th}$~OFDM symbol. For instance, the data symbols contained in $X_k[n]$ may be modulated using, \eg, binary phase shift keying~(BPSK) or quadrature amplitude modulation~(QAM) of arbitrary order. Then, in the time domain, a cyclic prefix 
is prepended to each OFDM symbol. At the receiver side~(see Fig.~\ref{fig:ofdm_receiver}), after time- and frequency synchronization, removal of the cyclic prefix, and discrete Fourier transform, the received baseband signal on the $k^{th}$~subcarrier of the $n^{th}$~OFDM~symbol in the frequency domain is given by:  
\begin{equation}
    Y_k[n] = H_k[n]\ X_k[n] + Z_k[n], \label{eq:tx_rx} 
\end{equation}
where $H_k[n]$ is the complex channel gain of the link between Alice and Bob for the $k^{th}$~subcarrier, and $Z_k[n] \sim \mathcal{CN}(0,\sigma^2)$ is additive white Gaussian noise~(AWGN). Following the implementation of practical systems, we assume that (known) pilot symbols are transmitted with a preamble to allow channel estimation at the receiver side. The pilot symbols are populated on each of the $K$~subcarriers of the $n^{th}$~OFDM~symbol (\ie, block-type pilot arrangement \cite{coleriChannelEstimationTechniques2002}) and allow Alice and Bob to obtain CSI using, \eg, a standard Least-Squares~(LS) channel estimator:
\begin{gather}
    \hat{H}_{k}[n] = \frac{Y_k[n]}{X_k[n]} = H_k[n] + \frac{Z_k[n]}{X_k[n]} =  H_k[n] + \tilde{Z}_k[n]. \label{eq:channel_probing_alice}
\end{gather}
The channel estimate then is used to equalize the subsequently received OFDM symbols:
\begin{equation}
    \hat{X}_k[n] = \frac{Y_k[n]}{\hat{H}_{k}[n]} 
    \label{eq:LS_equalization}
\end{equation}

\subsubsection{Intelligent Reflecting Surface}
We now establish the model for OFDM wireless communication in the presence of an IRS. We assume an IRS consisting of $N$ identical sub-wavelength-sized elements, arranged in an array on a planar surface to reflect impinging waves with a programmable phase shift. The generalized reflection coefficient for the $i^{th}$~IRS~element can be expressed as:
\begin{equation}
    r_i = \alpha_i e^{j \phi_i} \qquad i = 1,...,N,
    \label{eq:reflection_coeff}
\end{equation}
where we assume $\alpha_i = 1$ and $\phi_i \in [0, 2 \pi)$. Note that the IRS used in the experiments in Section~\ref{sec:experiments} is a binary phase-tunable IRS, \ie, then $\phi_i \in \{0, \pi\}$ and $r_i \in \{-1, 1\}$ which correspond to '0' and '1'~states of the IRS control signal. Next, following the illustration in Fig.~\ref{fig:system_model}, we find an expression for the channel between Alice and Bob, taking the IRS contribution into account. Here we assume that the non-IRS channel is static and therefore denote the IRS as only source of channel variation depending on $n$. The effective channel between Alice and Bob in (\ref{eq:tx_rx}) then is:
\begin{equation}
    H_k[n] = H_k^{IRS}[n] + d_k = \sum_{i=1}^N c_{i,k}\, r_i[n]\, g_{i,k} + d_k,
    \label{eq:h_eff}
\end{equation}
where $c_{i,k}, g_{i,k}, d_k \in \mathbb{C}$, respectively, are the complex channel gains of the link between Alice and the $i^{th}$~IRS~element, Bob and the $i^{th}$~IRS~element, the direct link between Alice and Bob for the $k^{th}$ subcarrier~(\cf~Fig.~\ref{fig:system_model}).

\subsection{Analytical Analysis}
We now proceed to show how the fast channel variations invoked by the ERA will impact OFDM wireless communication.

\subsubsection{Channel Equalization}
A fundamental part of every OFDM receiver (\cf~Fig.~\ref{fig:ofdm_receiver}) is the channel estimation that is mandatory to equalize the received data symbols~\cite{chiuehBasebandReceiverDesign2012}. As previously outlined, operating an IRS allows the attacker to alter the wireless channel between Alice and Bob which will thus likewise affect the channel equalization.

We assume the non-IRS channel $d_k$ is static and Eve switches between two IRS configurations $r^{(0)}_{i}$ and $r^{(1)}_{i}$, corresponding to the channels $H^{(0)}_{k}$ and $H^{(1)}_{k}$. 
Now consider the pilot symbols for channel estimation have been transmitted with the malicious IRS configured as $r^{(0)}_{i}$. Using (\ref{eq:channel_probing_alice}), the victim receiver obtains the following channel estimate:
\begin{equation}
    \hat{H}_k[n] = H^{(0)}_{k} + \tilde{Z}_k[n].
\end{equation}
Now, Eve switches the IRS configuration to $r^{(1)}_{i}$, changing the channel of the subsequent OFDM symbols to $H^{(1)}_{k}$. Thus, the victim receiver's equalizer, \cf (\ref{eq:LS_equalization}), will operate with an outdated channel estimation:
\begin{equation}
    \hat{X}_k[n] = \frac{Y_k[n]}{\hat{H}_k[n]} = \frac{X_k[n]\ H^{(1)}_{k} + Z_k[n]}{H^{(0)}_{k} + \tilde{Z}_k[n]},
\end{equation}
leading to a symbol error of
\begin{align}
    e_k[n] &= \hat{X}_k[n] - X_k[n] \nonumber\\
    &= \frac{X_k[n] \left( H^{(1)}_{k} - H^{(0)}_{k} - \tilde{Z}_k[n] \right) + Z_k[n]}{H^{(0)}_{k} + \tilde{Z}_k[n]}. \label{eq:symbol_err_without_noise}
\end{align}
For high SNRs, which is a reasonable assumption when using LS channel estimation, the symbol error is approximated by
\begin{equation}
    e_k[n] \approx X_k[n] \frac{ H^{(1)}_{k} - H^{(0)}_{k} }{H^{(0)}_{k}} = X_k[n] \frac{ H^{IRS,(1)}_{k} - H^{IRS,(0)}_{k}}{ H^{IRS,(0)}_{k} + d_k } 
    \label{eq:symbol_err_with_noise}
\end{equation}
The resulting expression in (\ref{eq:symbol_err_with_noise}) tells us that the IRS-induced symbol error is proportional to ($i$)~the transmitted symbol, ($ii$)~the difference between the two IRS channels, and ($iii$)~is inversely proportional to the direct channel contribution. Thus, the attacker can maximize its chance of causing a false symbol decision by producing a pair of IRS channels, \eg, ${H^{IRS,(1)}_{k} = -H^{IRS,(0)}_{k}}$. In particular, this can be achieved by inverting the sign of all IRS reflection coefficients $r_i$. Thus, we likewise adopt this approach in our simulations and experiments in Sections~\ref{sec:sim_results}~and~\ref{sec:experiments}.

\subsubsection{Intercarrier Interference}
OFDM systems in general are susceptible inter-carrier interference~(ICI) which is caused by a degradation of subcarrier orthogonality. ICI usually results from imperfections such as Doppler shifts, frequency offsets, and channel variations during an OFDM symbol period~\cite{goldsmith_wireless_2005, chiuehBasebandReceiverDesign2012}. We emphasize that the time-varying IRS used in the ERA will deliberately introduce rapid and instantaneous channel variations at sub-symbol timing resulting in substantial ICI. To model the ICI, (\ref{eq:tx_rx}) is modified to account for the interference  $H_{k,k'}$ from other subcarriers $k' \neq k$ to the received OFDM signal on the $k^{th}$ subcarrier~\cite{chiuehBasebandReceiverDesign2012}:
\begin{equation}
    Y_{k}[n] = H_{k}[n] X_{k}[n] + \underbrace{\sum_{k' \neq k}^{} H_{k,k'}[n] X_{k'}[n]}_{\textrm{ICI}}\ +\ Z_{k}[n].
\end{equation}
In  Appendix~\ref{appendix:ici} we show that if the ERA-induced fast channel variations are zero-mean over one OFDM symbol, the signal-to-interference ratio~(SIR) on the $k^{th}$ subcarrier is given by
\begin{equation}
    SIR_k =  \frac{S_k}{I_{IRS}}= \frac{|d_{k}|^2}{P_{IRS}},
\end{equation}
which means that the IRS does not contribute to the direct signal power $S_k$, but the total power received from the
IRS, $P_{IRS}$, completely translates into ICI, $I_{IRS}$, only.
Most importantly, this result is valid even without any optimization of the IRS elements with respect to the channels of the legitimate parties.

\section{Simulation Results}
\label{sec:sim_results}
After having analytically outlined the key mechanisms of the ERA affecting an OFDM system, we now strive to further explore the attack through simulations. We give comprehensive results, identifying attack parameters, including signal power, attacker distance, and IRS dimensions. Further, we show that the ERA leads to significant packet error rates~(PER) and is way more efficient when compared with a classical jamming attack using noise signals. 

As an example for general OFDM-based radio systems, we consider Wi-Fi here, since our experimental investigation following in Section~\ref{sec:experiments} also builds upon Wi-Fi devices. As the underlying simulation environment, we choose the MATLAB WLAN toolbox~\cite{wlan_tb_website} due to the availability of end-to-end simulation capabilities for the entire IEEE 802.11n physical layer, including channel coding and standard-compliant channel models. We summarize the essential simulation parameters in Table~\ref{tab:sim_params}. 
To mimic the adversarial IRS operation in the ERA, we add time-varying reflection, \ie, a complex square wave signal from the IRS, to one tap of the CIR.
Further, we randomize the time instant of the packet start with respect to the IRS modulation. For fairness in comparing the error rates across different modulation and coding schemes~(MCS), we adjust the packet payload sizes to always result in $16$ entire OFDM data symbols, regardless of the MCS setting. \mbox{Wi-Fi} uses an OFDM symbol duration of $\SI{4}{\us}$ and thus, the data portion of transmitted packets has a duration of \SI{64}{\us}.

Like traditional jamming attacks, the ERA is subject to link budget constraints. Thus, the attack efficiency depends on the signal power arriving at the receiver from the attacker. Although in the ERA the attacker does not generate a jamming signal itself, we can still define a \textit{jamming-to-signal ratio}~(JSR) as the ratio of IRS signal to direct (non-IRS) signal powers
\begin{equation}
    JSR = \frac{P_{IRS}}{S}=\frac{P_{IRS}}{\sum_{k}S_k}.
\end{equation}
For our simulations below, we use the JSR to assess the attacker strength. As an indication for the attacker's success, we leverage the PER.

\begin{table}[ht]
\small
\centering
\caption{Summary of the simulation parameters}
\label{tab:sim_params}
\begin{tabular}{@{}rl@{}}
\toprule
Component    & Parameter\\
\toprule
Wireless standard & IEEE 802.11n \\
Mode & HT Mixed \\
Bandwidth & \SI{40}{\MHz} \\
MIMO channels & $1$ \\
MCS index & $0$~-~$7$ \\
Total packet duration & \SI{92}{\us} \\
Data symbol duration & \SI{64}{\us} \\
Channel Model & Model D \\
Equalizer & Zero forcing \\
\bottomrule 
\end{tabular}
\end{table}

\subsection{Attacker Signal Power}
\label{sec:sim_jsr}

\begin{figure}
\includegraphics[width=0.95\linewidth]{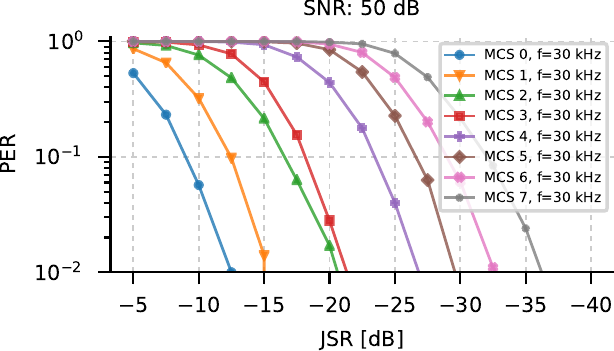}
\caption{End-to-end PER simulation results for IEEE 802.11n Wi-Fi under an ERA with \SI{30}{kHz} over varying JSRs for various modulation and coding schemes.}
\label{fig:sim_per_jsr}
\end{figure}

We investigate the victim PER performance as a function of the JSR for various MCS settings. Therefore, we assume the attacker signal originating from the IRS to have constant power while periodically toggling the phase between $0$ and $\pi$ at a rate of \SI{30}{kHz}, as is the case when inverting the sign of all IRS reflection coefficients $r_i$. The legitimate receiver has a high SNR of \SI{50}{dB}. We plot the PER results for MCS $0$~-~$7$ (covering BPSK, QPSK, 16-QAM, and 64-QAM modulations on the subcarriers~\cite{mcs_index_website}) as a function of the JSR in Fig.~\ref{fig:sim_per_jsr}. As expected, higher order modulations are more prone to interference from an ERA. The results also highlight that the ERA indeed is capable of producing error rates which render reliable wireless communication impractical.

\begin{figure}
\includegraphics[width=0.95\linewidth]{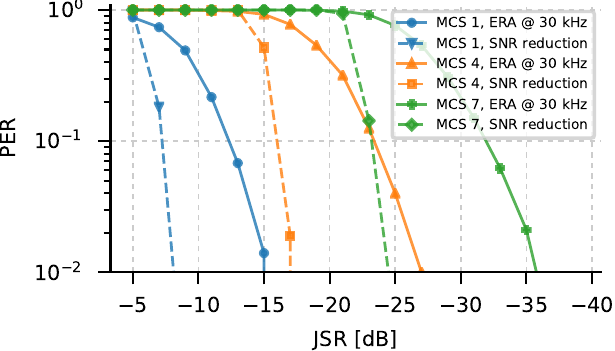}
\caption{End-to-end PER simulation results for IEEE 802.11n Wi-Fi to compare an ERA against SNR reduction, \eg, from noise jamming or signal power reduction. For the ERA case, we assume a noise-free channel.}
\label{fig:sim_era_vs_snr_reduction}
\end{figure}

To relate the ERA performance to classical noise-based jamming or signal power reduction attacks~\cite{lyuIRSBasedWirelessJamming2020}, we compare the attack against an SNR reduction. For the ERA, we now consider the legitimate receiver to have an otherwise noise-free channel. For the SNR reduction, we consider the IRS to remain static while the attacker now deteriorates the SNR by adding noise with power equivalent to the IRS signal strength during the ERA. 
We plot the PER simulation results in Fig.~\ref{fig:sim_era_vs_snr_reduction}, which indicates that the ERA achieves considerably better jamming performance when compared to a noise jammer at the same power.

\subsection{Channel Modulation Frequency}
\label{sec:sim_freq_per}
\begin{figure}
\includegraphics[width=0.95\linewidth]{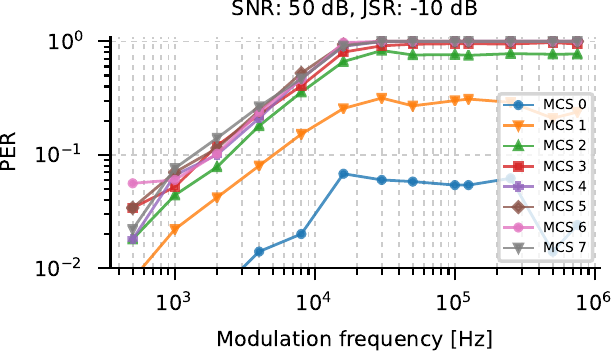}
\caption{End-to-end PER simulation results for IEEE 802.11n Wi-Fi for the ERA over channel modulation frequency for varying modulation and coding schemes at an SNR of \SI{50}{dB} with JSR of \SI{-10}{dB}.}
\label{fig:sim_per_freq}
\end{figure}

To fully characterize the ERA, we vary the IRS modulation frequency. We conduct the simulation for MCS indicies $0$~-~$7$ at an SNR of \SI{50}{dB} for the channel between Alice and Bob and a JSR of \SI{-10}{dB}. We plot the PER simulation results in Fig.~\ref{fig:sim_per_freq} against the IRS update frequency. For the MCS indices $0$ and $1$, we observe particularly lower PERs due to the more robust modulation parameters. Despite that, the PER clearly increases as a function of the modulation frequency for all MCS values. The increasing PER at lower modulation frequencies can be explained by the increasing probability of an IRS reconfiguration taking place during packet transmission. That is, the packet error rate resulting from an ERA with IRS pattern durations $T_{IRS}$ longer than the packet duration $T_p$ is upper bounded by $T_p/T_{IRS}$. As the PER for modulation frequencies above approximately \SI{16}{kHz} reaches a plateau, we conclude that at least one IRS reconfiguration during transmission of the data symbols suffices to achieve the maximum attack efficiency for a certain JSR.

\subsection{Surface Size}

\begin{figure}
\centering
\subfloat[]{%
        \includegraphics[width=0.65\columnwidth]{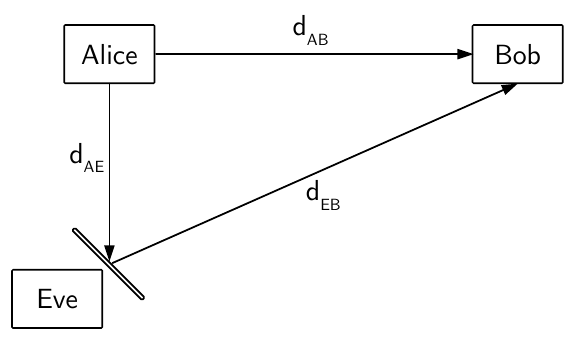}
        }
\hfill
\subfloat[]{%
        \includegraphics[width=1.0\columnwidth]{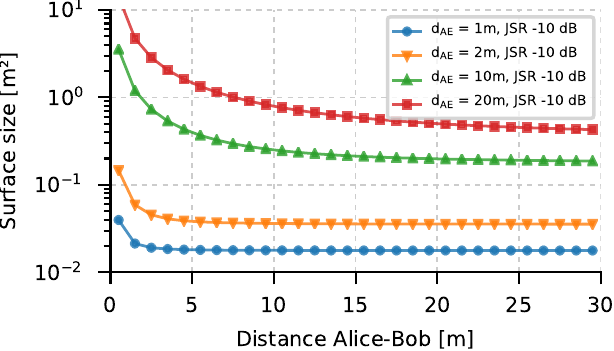}
        }
\hspace*{\fill}%
\caption{Simulation of the minimum surface size requirement for to achieve a JSR of \SI{-10}{dB}. (a)~Geometrical configuration used for the simulation, indicating the relative positions of Alice, Bob, and Eve's IRS. (b) Minimum IRS size versus $d_{AB}$ for varying attacker distances $d_{EA}$, assuming free-space path loss at~\SI{5.35}{GHz}.}
\label{fig:sim_surface_size}
\end{figure}

We will now show that an ERA is feasible even for rather weak attacker configurations regarding the attacker distance and IRS dimensions. Previously, we have determined the JSRs necessary for the attacker to degrade the PER of Alice and Bob (see Fig.~\ref{fig:sim_per_jsr}). Note that we define the JSR as the ratio of the signal power coming from the IRS and the direct (non-IRS) signal power. Thus, the attacker generally seeks to pick up sufficient power from the legitimate users. The attacker can either minimize the distance to one of the victim parties to minimize path loss or increase the IRS size. Although both strategies are suitable, we assume the attacker must maintain a minimum distance and also cannot increase the IRS size arbitrarily without raising suspicion. Hence, we derive a connection between JSR, attacker distance, and the surface size. For the parties, we assume the geometrical configuration shown in Fig.~\ref{fig:sim_surface_size}~(a). We start with the free-space path loss of the direct link between Alice and Bob~\cite{goldsmith_wireless_2005}, where the received power is proportional to
\begin{equation}
    L_d = \left( \frac{\lambda}{4 \pi d_{AB}} \right)^2,
\end{equation}
with the carrier frequency wavelength $\lambda = c_0/f$. For an optimal surface configuration, the free-space path gain from Alice to Bob via the IRS is found by~\cite{ozdoganIntelligentReflectingSurfaces2020}:
\begin{equation}
    L_{IRS} = \left( \frac{A_{IRS}}{4 \pi d_{AE} d_{EB}} \right)^2.
\end{equation}
Assuming Alice and Bob use omni-directional antennas, the JSR becomes
\begin{equation}
    JSR = \frac{L_{IRS}}{L_d} = \left( \frac{A_{IRS}\ d_{AB}}{d_{AE} d_{EB} \lambda} \right)^2,
\end{equation}
which allows us to link the surface area $A_{IRS}$ to the JSR:
\begin{equation}
    A_{IRS} = \sqrt{JSR} \frac{d_{AE} d_{EB} \lambda}{d_{AB}}
    \label{eq:jsr_area}
\end{equation}

We use Equation~(\ref{eq:jsr_area}) to plot the minimum IRS size required by an attacker to achieve a JSR of \SI{-10}{dB} in Fig.~\ref{fig:sim_surface_size}~(b). We show the result as a function of the distance between Alice and Bob and for distances \SI{1}{m}, \SI{2}{m}, \SI{10}{m}, and \SI{20}{m} of Eve to Alice. Consider, for example, Alice and Bob are at a distance of \SI{30}{\metre} and Eve is at a distance of \SI{10}{\metre} to Alice. Then, an IRS size of only \SI{0.19}{\square\metre} is sufficient to achieve a JSR of \SI{-10}{\dB}, which results in a severe PER degradation for Alice and Bob.

\section{Experimental Evaluation}
\label{sec:experiments}

After having approached the ERA through theoretical analysis and simulations in the previous sections, we now proceed with a practical evaluation of the ERA. Therefore, we first describe our experimental setup comprising of a low-cost IRS prototype and commodity \mbox{Wi-Fi} devices. Furthermore, we demonstrate that the ERA is capable of severe link quality degradation, leading to a significant reduction in the effective wireless data throughput.

\subsection{Experimental Attack Setup}
In this section, we present our experimental attack setup consisting of a prototype IRS and two microcontrollers. We estimate the cost of the setup to be around~\EUR{100}\footnote{\EUR{40} for microcontroller development boards, \EUR{30} for PCBs, \EUR{30} for surface-mount components.}. 

\subsubsection{IRS Prototype}

\begin{figure}
\centering
\hspace*{\fill}%
\subfloat[]{{%
        \includegraphics[width=0.59\columnwidth]{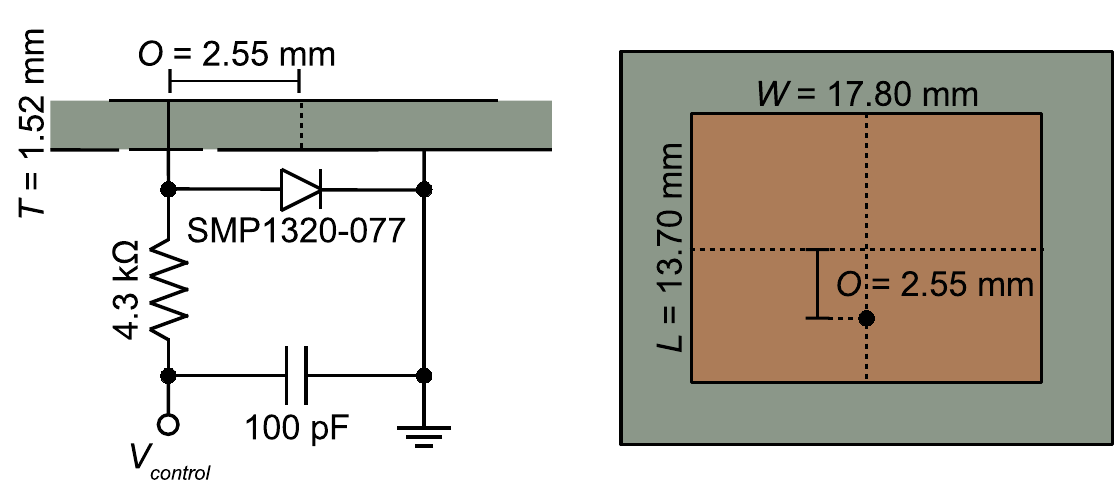}
        }}
\hfill
\subfloat[]{{%
        \includegraphics[width=0.39\columnwidth]{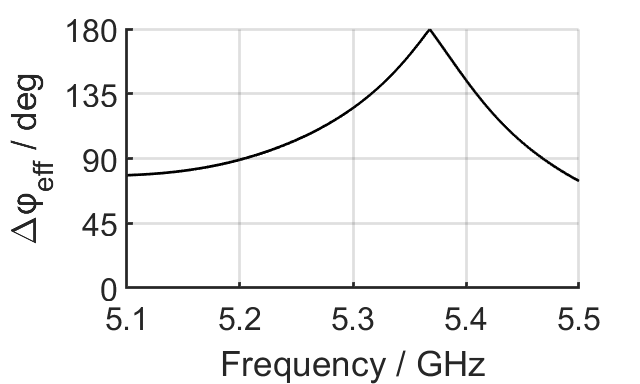}}}
\hspace*{\fill}%
\caption{(a) Unit cell schematic and dimensions. (b) Unit cell phase response over frequency.}
\label{fig:irs_unit_cell_phase}
\end{figure}

\begin{figure}
\centering
\hspace*{\fill}%
\subfloat[]{%
        \includegraphics[width=0.49\columnwidth]{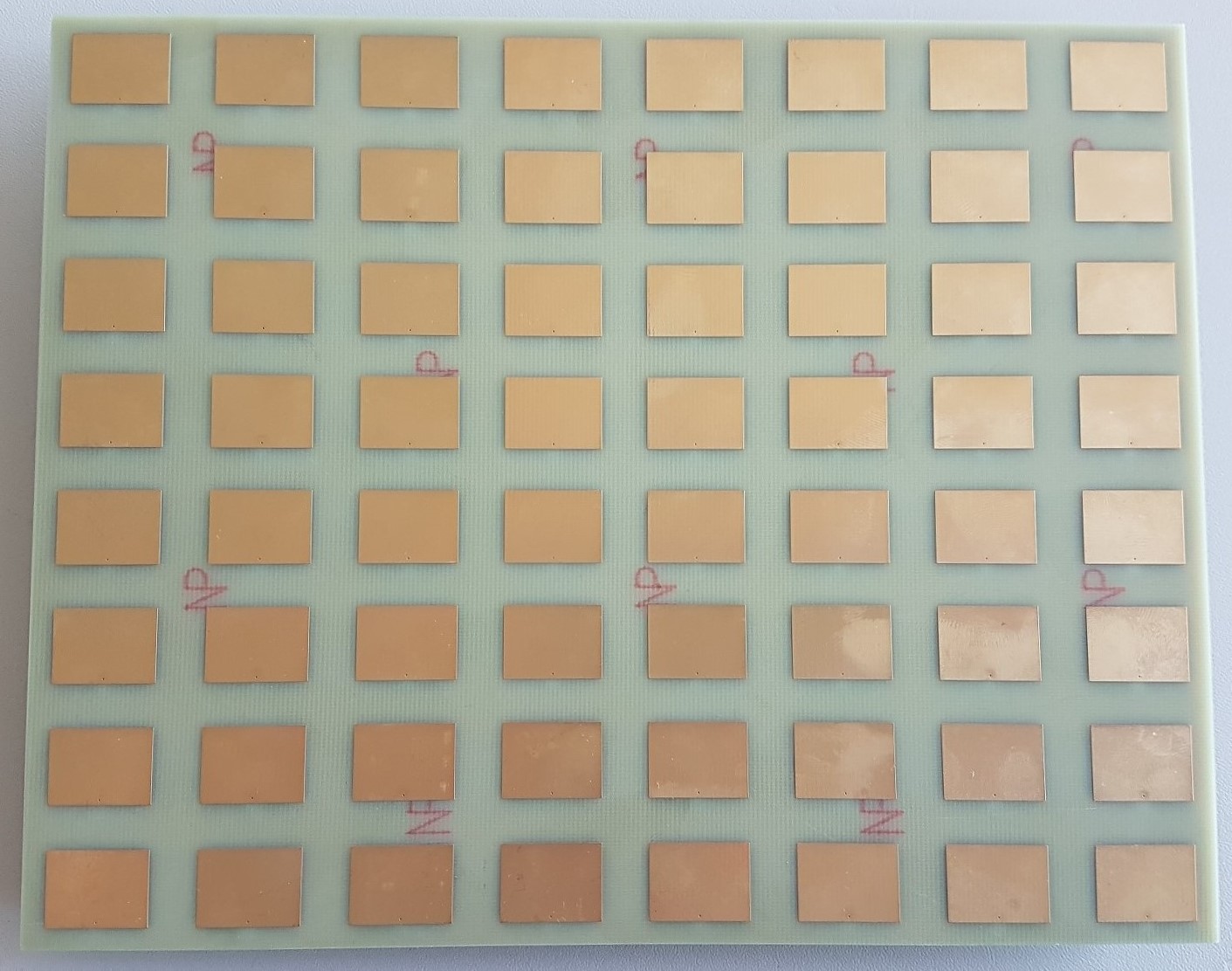}
        }
\hfill
\subfloat[]{%
        \includegraphics[width=0.49\columnwidth]{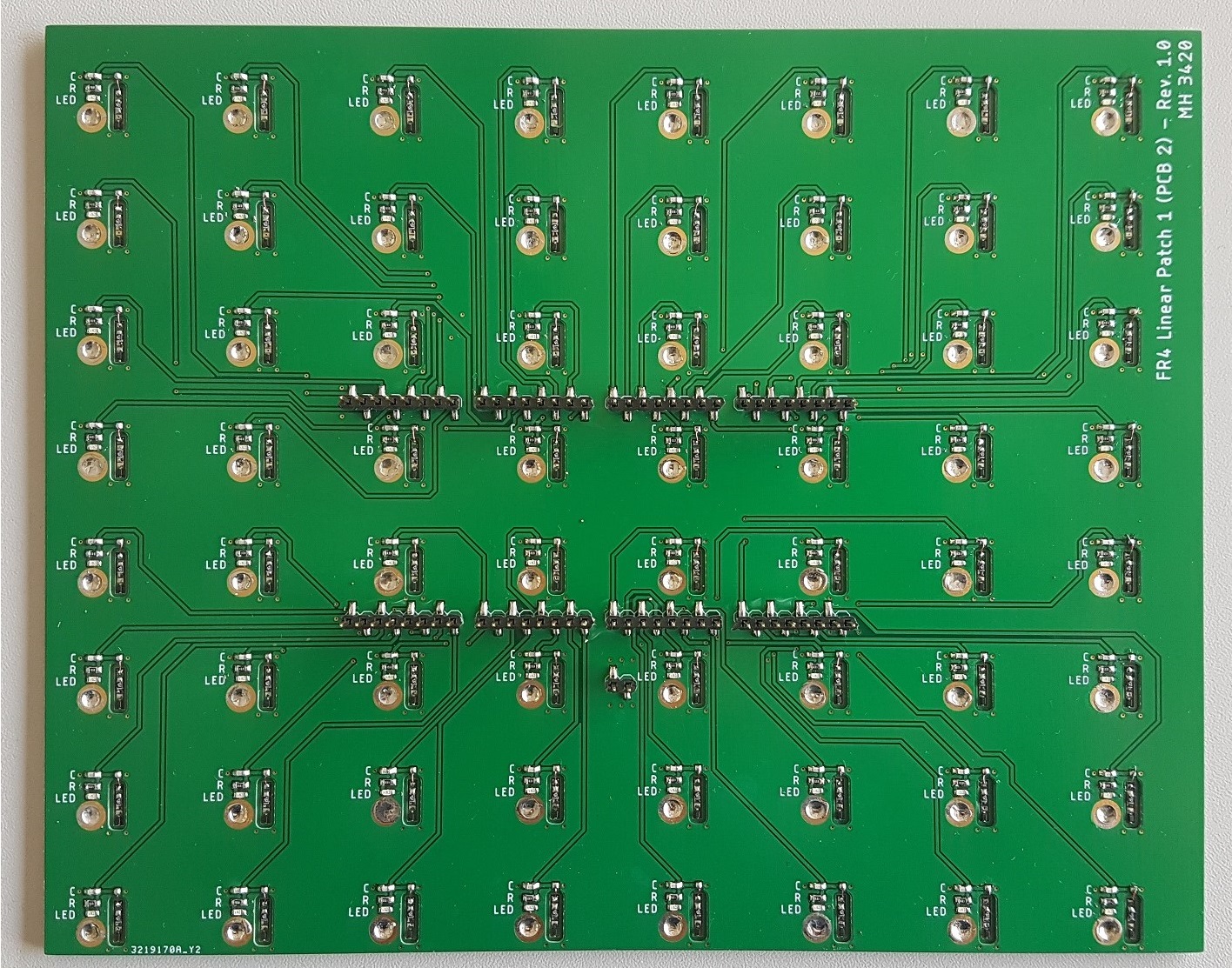}}
\hspace*{\fill}%
\caption{Intelligent reflecting surface prototype module. (a) Front view with patch elements (\SI{20}{cm} x \SI{16}{cm}). (b) Back view with control lines, PIN diodes, and biasing circuitry.}
\label{fig:irs_fotos}
\end{figure}

As the essential part of a first exploration of the ERA in practical experiments, we use two low-cost IRS prototype modules (see Fig.~\ref{fig:irs_fotos}~(a)) with $128$~binary-phase tunable unit-cell elements in total, arranged in a $16 \times 8$~array on standard FR4 PCB substrate. The elements are rectangular patch reflectors on top of a ground plane. Attached to each element, there is a PIN diode which can switch a parasitic element to the reflector, allowing to shift its resonance frequency. Thereby, the reflection coefficient of each element can be individually switched between two states, \ie, a '0'~state and a '1'~state, by turning the control voltage to the reflector element either on or off. The unit cell circuitry and the reflector design are shown in Fig.~\ref{fig:irs_unit_cell_phase}~(a). The IRS prototype used in our experiments is optimized to achieve a \SI{180}{\degree} phase difference in the reflected wave for the '0' and '1'~states (see Fig.~\ref{fig:irs_unit_cell_phase}~(b)), \ie, $r_i \in \{-1, 1\}$ in (\ref{eq:h_eff}). 

\subsubsection{IRS Modulation}
As we strive for rather high IRS modulation frequencies, we drive the $128$~IRS~elements in parallel. Therefore, we connect each of the $128$~control~lines to a GPIO pin of two STM32F407 microcontrollers, allowing us to achieve IRS modulation frequencies of up to \SI{1.6}{\MHz}. The frequency and surface patterns used for the modulation are programmable from the host controller through an UART serial communication interface.

Like in the theoretical analysis and the simulations, \cf~Section~\ref{sec:sim_results}, we apply a simple binary surface modulation. That is, we periodically toggle between two IRS configurations and thereby maintain a low attack complexity. For instance, we switch between all $128$~IRS~elements either set to the '0' or '1'~state. As discussed in Section~\ref{sec:analytical_analysis}, since $r_i \in \{-1, 1\}$, this leads to switching between two channels $H^{(0)}_{k}$ and $H^{(1)}_{k}$, with ${H^{IRS,(1)}_{k} = -H^{IRS,(0)}_{k}}$.

\subsection{Wireless Throughput Measurement}
\begin{figure}
\centering
\includegraphics[width=0.9\linewidth]{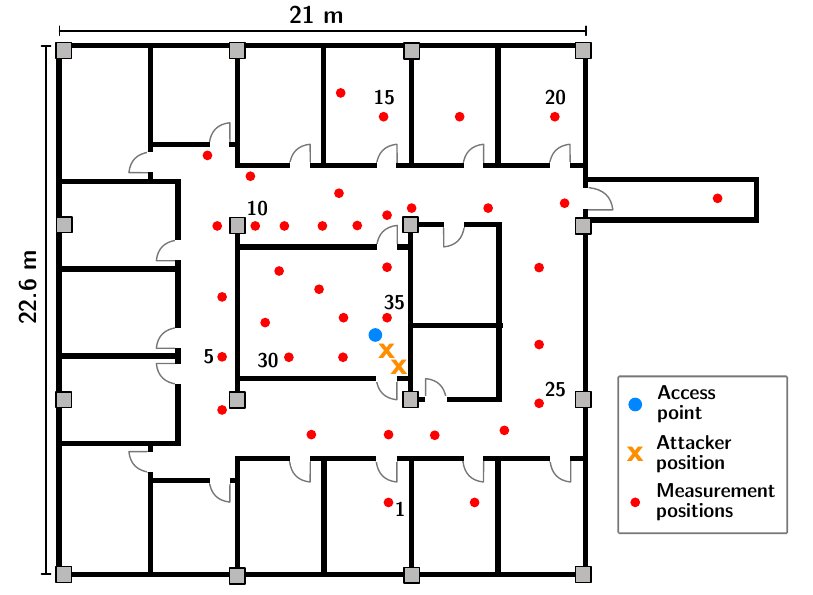}
\caption{Floorplan of the office space used for throughput measurements, indicating the positions of the WLAN router (access point), the attacker setup, as well as each of the $37$ throughput measurement positions.}
\label{fig:floorplan}
\end{figure}

\begin{figure}
\centering
\includegraphics[width=0.9\linewidth]{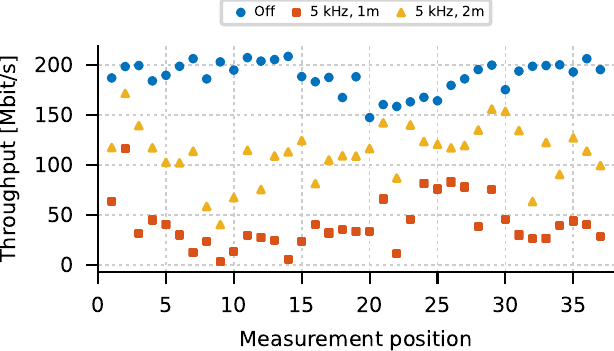}
\caption{Throughput measurement results from testing download speeds at $37$ positions in the office space with and without the ERA taking place.}
\label{fig:3og_results}
\end{figure}

We now demonstrate that the ERA is capable of significant throughput reduction in entire wireless networks. Therefore, we deploy a commercial off-the-shelf WLAN router to provide an IEEE 802.11ac network in an office space. We position the attacker setup strategically at the router with distances of \SI{1}{\metre} and \SI{2}{\metre}. We detail and summarize the setup in Table~\ref{tab:params}.

For the experiment, we use a laptop connected to the Internet via the \mbox{Wi-Fi} network to measure the effective end-to-end speed of the connection~\cite{speedtest_cli_git}. We perform speed measurements without the ERA (the malicious IRS remains static) and with the ERA enabled (switching all IRS elements between '0' or '1'~state). We repeat this procedure for a total of $37$~positions distributed throughout the office space, as indicated in Fig.~\ref{fig:floorplan}. 
We show the results of the throughput measurements in Fig.~\ref{fig:3og_results}. Here we can see that the ERA leads to an average throughput reduction of $78$~\% and $40$~\% for the attacker at \SI{1}{\metre} and \SI{2}{\metre} distance to the router, respectively. Recall that the attacker does not actively emit any jamming signal to achieve this result. Furthermore, the attacker does not perform any kind of synchronization to the legitimate signals or optimization of the IRS configurations. Notably, the ERA also leads to substantial throughput reduction where the wireless channel between the client and the IRS is obstructed, \ie, in different rooms with walls in between. Thus, we conclude that the ERA is a scalable attack, allowing the attacker to slow down the wireless network at many different places.

\begin{table}[ht]
\small
\centering
\caption{Summary of the experimental setup}
\label{tab:params}
\begin{tabular}{@{}rl@{}}
\toprule
Component    & Parameter\\
\toprule
\textbf{Jammer}  &  \\
Surface elements & 128\\
Surface size & \SI{40}{\centi\metre}~$\times$~\SI{16}{\centi\metre}, \SI{0.064}{\square\metre} \\
Operation frequency & \SI{5.37}{\GHz} \\
Modulation frequency &  \SI{5}{\kHz} \\
Modulation type & All '0' / all '1' states\\
\midrule
\textbf{Wi-Fi}  &  \\
Access point & Asus RT-AC59U V2 \\
Client & \begin{tabular}{@{}l@{}}Dell Latitude 7490 Laptop,\\Intel Wireless-AC 8265\end{tabular} \\
Standard & IEEE 802.11n/ac \\
Frequency & Channel 64, \SI{5.32}{\GHz} \\
Bandwidth & \SI{40}{\MHz} \\
MIMO channels & 2 \\
\bottomrule 
\end{tabular}
\end{table}

\begin{figure}
\centering
\includegraphics[width=0.9\linewidth]{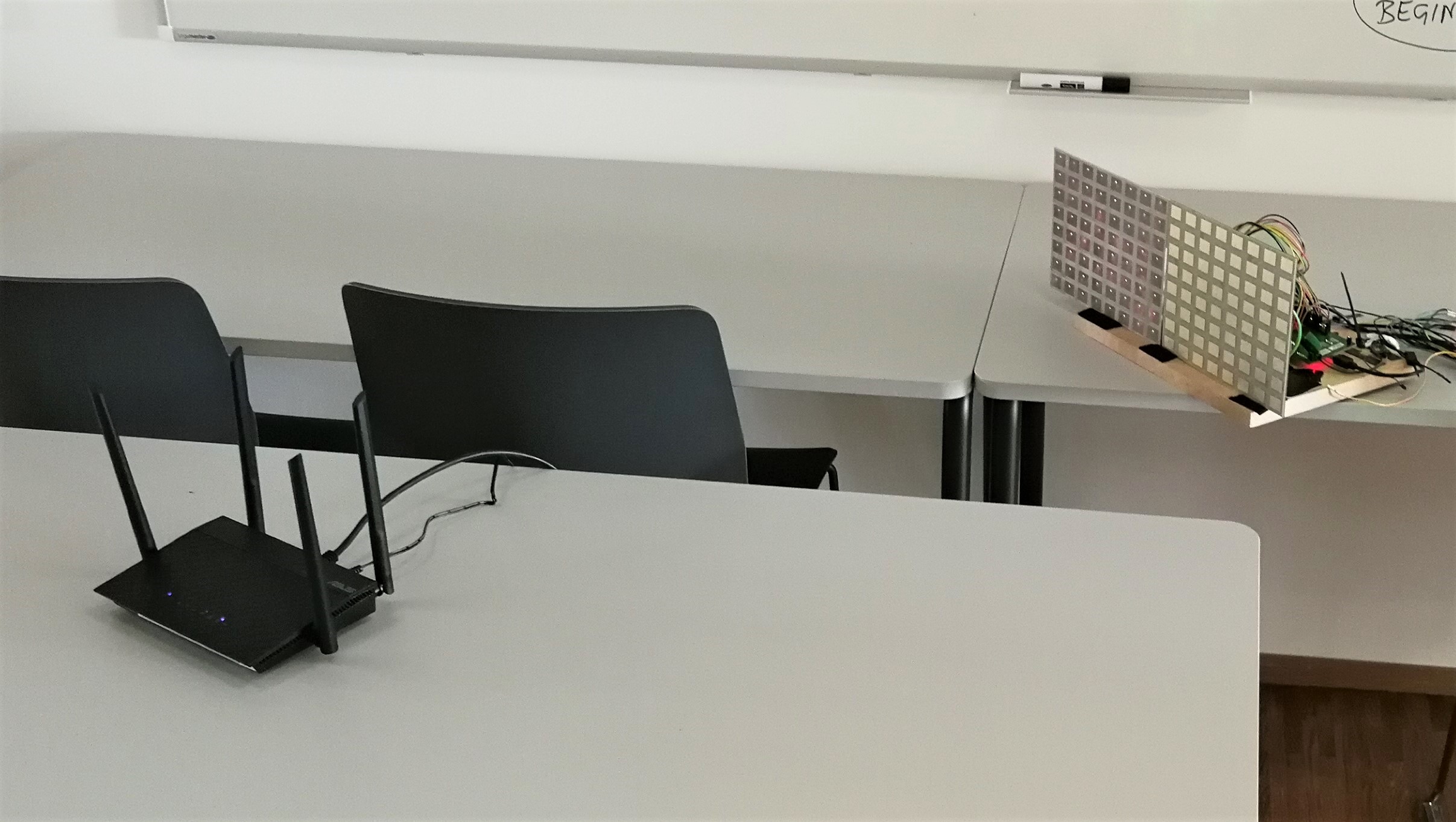}
\caption{Experimental ERA setup with WLAN router and attacker IRS.}
\label{fig:setup_foto}
\end{figure}

\subsection{Systematic Packet Error Rate Measurement}
We perform a second experiment to systematically assess the practical effectiveness of the ERA, aiming to obtain PER measurements similarly to our simulation result from Section~\ref{sec:sim_freq_per}. Therefore, we deploy single-board computers equipped with ath9k-based network interface cards~(NICs)~\cite{xie_precise_2015} for IEEE 802.11n \mbox{Wi-Fi} at the legitimate parties Alice and Bob. The NICs give us low level access to the \mbox{Wi-Fi} communication, \ie, we can transmit packets with defined length and MCS setting. Here, we use a 2x2 MIMO configuration with off-the-shelf \mbox{Wi-Fi} antennas. One of the parties provides a \mbox{Wi-Fi} network on channel~$60$ (at \SI{5300}{\MHz}), allocating \SI{40}{\MHz} bandwidth. We place the attacker setup attacker at distance \SI{2}{\metre} and \SI{3}{\metre} in line-of-sight to Alice and Bob, respectively. The channel between Alice and Bob also has line-of-sight conditions. For the whole duration of the experiment, the propagation environment remains static apart from the adversarial IRS operation.

\begin{figure}
\centering
\subfloat[]{%
        \includegraphics[width=0.95\columnwidth]{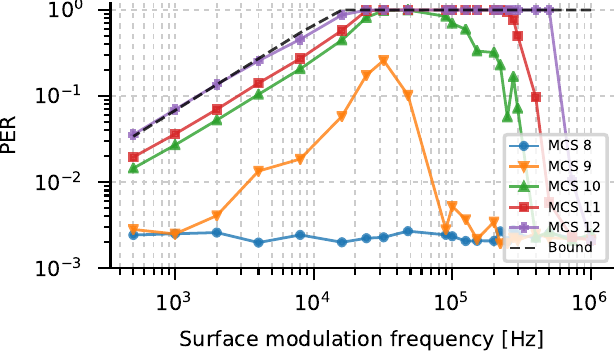}
        }
\hfill
\subfloat[]{%
        \includegraphics[width=0.95\columnwidth]{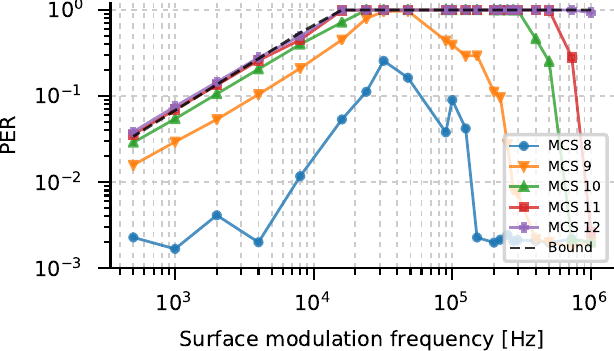}}
\caption{Measured PER over channel modulation frequency. (a) Binary pattern modulation. (b) Tailored pattern modulation.}
\label{fig:per_vs_freq}
\end{figure}

In our setup, Alice transmits $20000$~packets with randomized payload data to Bob. For each transmission, we configure the payload size and the MCS setting. Similarly to the simulation, we adjust the payload size to always result in $9$~entire OFDM symbols (data symbol duration \SI{3.6}{\us}, packet duration \SI{6.8}{\us}). On Bob's side, we count the number of successfully received packets to finally obtain the PER. We plot the PER results as a function of the adversarial IRS modulation frequency in Fig.~\ref{fig:per_vs_freq}~(a). Also, we indicate the previously discussed upper PER bound given by $T_p/T_{IRS}$ for $T_{IRS} > T_p$. Essentially, our measurement with standard Wi-Fi NICs confirms our previous simulation results, showing that higher-order modulations are more susceptible to the ERA. However, instead of reaching a plateau, we observe a drop in the PER when increasing the IRS modulation frequency beyond \SI{30}{kHz}. We believe that this effect is due to hardware imperfections on the IRS prototype which initially was not designed to operate at high modulation speeds. As evident from the results, the upper PER bound based on the timing parameters holds. However, despite the fixed packet time duration, it appears that our bound seems to be too optimistic for MCS values below $12$. We attribute this to reduced synchronization efforts, \ie, the receiver will barely be affected by an IRS change during the packet's preamble portion, reducing the effective ERA-sensitive packet length.

\subsubsection{Surface Pattern Optimization}
\label{sec:targeted_pattern}
Thus far, we have tested the simplest ERA strategy where the attacker switches all surface elements periodically between the '0' or '1'~states. However, this strategy can be further improved by matching the used IRS configurations to the wireless link under attack. Thus, the attacker may prepend its jamming operation with a setup phase in order to optimize the IRS configurations used during the subsequent ERA. The attacker therefore can incorporate eavesdropped CSI feedback of the victim parties to further enhance the attack efficiency. For a first demonstration, we design and test an adaptive optimization algorithm to find IRS configurations well-suited for the ERA. The intuition of the algorithm is to use the adversarial IRS for maximizing a dissimilarity measure between the pair of IRS-induced channel responses of the victim wireless link. Following our analytical analysis in Section~\ref{sec:analytical_analysis}, we expect this to improve the attacker's success.
Algorithm~\ref{alg:surface_optimization} outlines the procedure. The result are two IRS configurations $r^{(0)}_{i}$ and $r^{(1)}_{i}$. Note that we here denote the binary surface control settings ('0' or '1') as a proxy for reflection coefficients. 

\begin{algorithm}[ht]
\SetAlgoLined
\KwResult{Distinct IRS configurations $r^{(0)}_{i}$, $r^{(1)}_{i}$ for ERA.}
 start with random $N$-bit IRS configurations $r^{(0)}_{i}, r^{(1)}_{i}$\;
 dissimilarity metric $d$\;
 algorithm rounds $R=2$\;
 \For{$j = 0$ to $R$}{
    configure IRS as $r^{(1)}_{i}$\;
    $ref^{(1)} \gets H_k(r^{(1)}_{i})$\;
    configure IRS as $r^{(0)}_{i}$\;
    \For{$i \gets 0$ to $N$}{
      $ref^{(0)}_{i,0} \gets H_k(r^{(0)}_{i})$\;
      $r^{(0)}_i \gets r^{(0)}_i \oplus 1$\;
      update IRS element $i$\;
      $ref^{(0)}_{i,1} \gets H_k(r^{(0)}_{i})$\;
      \If{$d(\textrm{ref}^{(1)}, ref^{(0)}_{i,0}) > d(\textrm{ref}^{(1)}, ref^{(0)}_{i,1})$}{
       $r^{(0)}_i \gets r^{(0)}_i \oplus 1$\;
       update IRS element $i$\;
       }
    }
    swap($r^{(0)}_{i}$, $r^{(1)}_{i}$)\;
 }
 \caption{Adversarial binary surface optimization}
 \label{alg:surface_optimization}
\end{algorithm}

The randomly chosen initial IRS configurations in Algorithm~\ref{alg:surface_optimization} are given below:\\
\makebox[\linewidth]{$r^{(0)}_{i}$ = \texttt{0x5CC81D86E5DAB902B071665D1D7DC2F1}}\\
\makebox[\linewidth]{$r^{(1)}_{i}$ = \texttt{0xC859CCA60594481B193BF3D236E877AE}}
The result of the algorithm are the updated IRS configurations:\\
\makebox[\linewidth]{$r^{(0)}_{i}$ = \texttt{0xFFFF9F9F08089E08474721D92AC1B57A}}\\
\makebox[\linewidth]{$r^{(1)}_{i}$ = \texttt{0x00006060E5D776A2F8B876020C034C05}}

\begin{figure}
\centering
\includegraphics[width=0.9\linewidth]{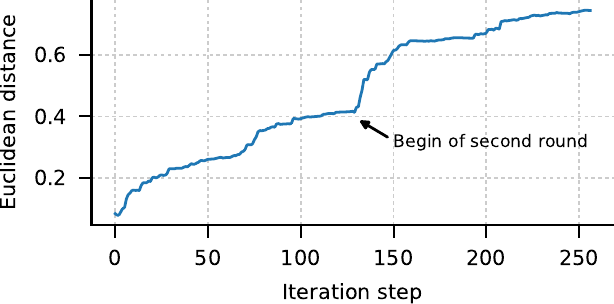}
\caption{Evolution of Euclidean distance between the channel responses during the iterative IRS optimization.}
\label{fig:ed_evolution}
\end{figure}

\begin{figure}
\centering
\includegraphics[width=0.9\linewidth]{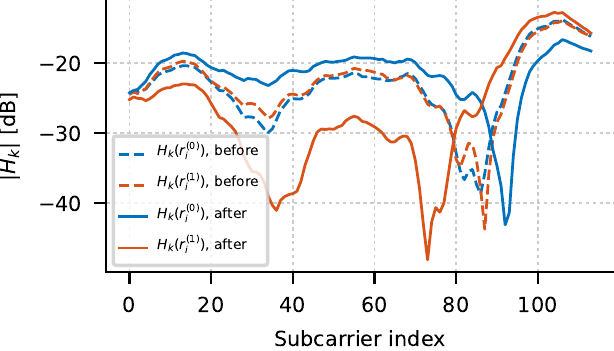}
\caption{Effective normalized channel responses observed by Alice and Bob, before and after running the adversarial IRS optimization algorithm.}
\label{fig:csi_results}
\end{figure}

Fig.~\ref{fig:ed_evolution} shows the evolution of the Euclidean distance between $|H_k(r^{(0)}_{i})|$ and $|H_k(r^{(1)}_{i})|$ over the iteration steps, clearly exhibiting the characteristic behaviour of our algorithm. Finally, we also plot the pair of channel responses as observed by Alice and Bob before and after the optimization in Fig.~\ref{fig:csi_results}. Here, we can see that our procedure indeed is highly effective in providing distinct channel responses designated to be used in the ERA. Note that even though the reception for $|H_k(r^{(0)}_{i})|$ has improved after running the algorithm, the difference between the two channel states is maximized. The result is a vivid example for the combination of inherent simplicity and possibilities of the IRS for previously infeasible attacks.

Using the presented algorithm with the Euclidean distance as a metric and magnitude CSI information on the link between Alice and Bob, we obtain the adapted IRS configurations $r^{(0)}_{i}$ and $r^{(1)}_{i}$, which we now use to conduct the ERA. We repeat the PER measurement experiment from the previous section and plot the results in Fig.~\ref{fig:per_vs_freq}~(b). Here it is evident that the optimization was able to improve the attacker efficiency. Now, even the robust BPSK modulation for MCS~$8$ exhibits a significant PER induced by the ERA. Further, the optimization has also led to substantially increased PERs for the remaining MCS values.

\section{Discussion}
\label{sec:discussion}
In this section, we discuss ($i$)~the real-world applicability, ($ii$)~the attacker capabilities, and ($iii$)~reason about countermeasures and mitigation. Also, we give directions for future work.

\subsection{Real-world Applicability}

We assess the costs and complexity of an ERA to be low. Our results show that a sub \EUR{100} attacker setup can have significant impact on the effective wireless throughput. Once an attacker possesses a functional IRS, only basic microcontroller programming is required to rapidly vary a number of logic signals controlling the IRS. Thus, the attack can be easily carried out by non-specialists. While the commercial availability of IRS devices is currently still limited, several companies~\cite{greenerwave_website, metawave_website} are working on product-grade IRS implementations. Besides that, many IRS designs are publicly available and can easily be reproduced by others using cheap PCB assemblies. Instead of using an own IRS, an attacker could also hijack existing IRS infrastructure which may be deployed in future wireless networks~\cite{yang6GWirelessCommunications2019}, most likely already at strategically advantageous positions.

\subsection{Attacker Capabilities}

To conduct an ERA, the attacker's IRS must be within the wireless propagation environment between the victim nodes. As wireless communication is inherently supposed to bridge distances this will not be a hurdle for an attacker. As discussed, the JSR is an important parameter bounding the attack performance. In order to improve its JSR, the attacker can choose a favorable position or increase the IRS size. 
Therefore, to compensate the small size of our IRS prototype, we have used rather short attacker distances in our experiments, which still represents a valid attacker model. Our simulation results show that sufficient JSR values are, in principle, still possible for higher attacker distances and surface sizes. However, this also reveals a limitation of ERA: the attacker is passive and cannot amplify the signals it reflects. Hence, as it is generally the case for wireless communications (and jamming), the attack is limited by the available link budget.

Our simulation results show the underlying relationship between JSR and PER. For this purpose, we have simplified the attacker's signal originating from the IRS to a time-varying signal component from alternating the sign of the IRS reflection coefficients. Although finding a corresponding IRS configuration to meet a certain JSR is non-trivial, our practical tests tests show that even with a binary-phase tunable IRS and without optimized surface configurations, the ERA significantly disrupts the victim communication.

In Section~\ref{sec:targeted_pattern}, we have granted the attacker access to the CSI of Alice and Bob to demonstrate that an attacker can further optimize the IRS configurations used during the ERA. In an actual attack, the attacker would rely on eavesdropping CSI feedback, \eg, from the user to the base station. For instance, this is commonly used in IEEE 802.11 WLAN standards, 4G, and 5G to implement, \eg, transmit beamforming~\cite{girkeResilient5GLessons2019, rothLocalizationAttackPrecoder2020, IEEEStandardInformation, etsiETSITS1382018}. Note that, in the standards mentioned, these signals are not encrypted.

\subsection{Countermeasures} The ERA is based on an IRS within the channel between Alice and Bob. For the attack to work, a part of the transmitted signal must reach the receiver via the adversarial IRS. Due to the broadcast nature of wireless signal propagation, it is likely that an ERA cannot generally be prevented. The transmitter could use beamforming to diminish the attacker's success, trying to minimize the signal power reaching the IRS. However, this requires a mechanism for attack detection and localization and an advanced attacker may even leverage beamforming to its favor by providing a preferred path via the IRS to the receiver. Since the interference signal produced in the ERA is correlated to the useful signal, it may also be possible to find signal processing-based countermeasures at the receiver side. However, we emphasize these considerations are speculative. Countermeasures, if they exist, cannot be implemented immediately in end-user equipment because the very low-level signal processing of radio transceivers is usually implemented in hardware or is not updatable.

Finally, to mitigate the attack, wireless communication systems could apply encryption of physical layer control channels, \ie, to prevent the attacker to obtain CSI feedback. However, this will not render the ERA infeasible, but would only impede an adversarial IRS optimization. Moreover, this requires drastic changes to protocols and such measures can likely only be implemented within future standards.

\subsection{Future work}
In this paper, we have presented a novel class of jamming attacks based on IRS-induced fast changes in the radio propagation environment of wireless communication parties. Naturally, this work only represents a very first exploration of the ERA and, more broadly, the IRS as a toolkit for practical wireless physical layer attacks. Therefore, our work may serve as a basis for future work studying, for example, the following aspects.

\Paragraph{Improving the attack} We have provided first insights into the optimization of the IRS configuration for an ERA, demonstrating the potential for increased attack efficiency. The evaluation of improved optimization algorithms based on eavesdropping CSI feedback is left for future work. Also, future work should investigate non-binary surface modulation signals where the attacker uses more than two IRS configurations. Finally, there is room for hardware improvements to the attacker setup, perhaps through dedicated IRS designs for high modulation frequencies.

\Paragraph{Attack detection and countermeasures} More work is needed to examine whether existing jamming attack detection and mitigation strategies, \eg, \cite{groverJammingAntijammingTechniques2014}, can be adapted to the ERA. Also, we see a need to evaluate the possibility of signal processing based mitigation strategies that could be incorporated into future transmitter and receiver architectures.

\Paragraph{Application to other modulations} We have outlined the ERA against OFDM communications, as it is the preferred modulation scheme for modern wireless communication systems, including Wi-Fi, 4G, 5G. Further studies should investigate the applicability of ERA to other modulation schemes.

\section{Conclusion}
\label{sec:conclusion}

In this paper, we have first used the IRS as a cost-effective attacker tool to accomplish physical layer attacks in wireless radio networks. Based on this observation, we introduce the Environment Reconfiguration Attack~(ERA) as a novel wireless jamming attack primitive. Without actively emitting a jamming signal, the ERA allows an attacker to significantly reduce or even disable the wireless communication capabilities of victim parties. Our approach takes advantage of a time-varying IRS which we use to rapidly modulate the channel response of victim wireless communication parties. Using the widespread OFDM modulation as an example, we have shown that exceptionally fast and instantaneous changes in the radio propagation environment disturb radio receivers substantially. We have approached the ERA through analytical analysis, simulations, and experiments. Our work breaks down the fundamental attack mechanisms and determines important attacker requirements before demonstrating multiple experimental attacks on actual wireless networks.

Our work highlights that the IRS must be considered as a powerful attacker tool for physical layer attacks against wireless communications. The IRS is a striking example of how emerging technologies are causing attack taxonomies to shift as previously complex attacks become tractable.

\section*{Acknowledgements}
This work was supported in part by the German Federal Ministry of Education and Research~(BMBF) within the project MetaSEC (Grant 16KIS1234K) and by the German Research Foundation~(DFG) within the framework of the Excellence Strategy of the Federal Government and the States - EXC2092 CASA - 390781972.

\bibliographystyle{ACM-Reference-Format}
\bibliography{passjam-refs}


\begin{thebibliography}{49}


\ifx \showCODEN    \undefined \def \showCODEN     #1{\unskip}     \fi
\ifx \showDOI      \undefined \def \showDOI       #1{#1}\fi
\ifx \showISBNx    \undefined \def \showISBNx     #1{\unskip}     \fi
\ifx \showISBNxiii \undefined \def \showISBNxiii  #1{\unskip}     \fi
\ifx \showISSN     \undefined \def \showISSN      #1{\unskip}     \fi
\ifx \showLCCN     \undefined \def \showLCCN      #1{\unskip}     \fi
\ifx \shownote     \undefined \def \shownote      #1{#1}          \fi
\ifx \showarticletitle \undefined \def \showarticletitle #1{#1}   \fi
\ifx \showURL      \undefined \def \showURL       {\relax}        \fi
\providecommand\bibfield[2]{#2}
\providecommand\bibinfo[2]{#2}
\providecommand\natexlab[1]{#1}
\providecommand\showeprint[2][]{arXiv:#2}

\bibitem[\protect\citeauthoryear{Agiwal, Roy, and Saxena}{Agiwal
  et~al\mbox{.}}{2016}]%
        {agiwalNextGeneration5G23}
\bibfield{author}{\bibinfo{person}{Mamta Agiwal}, \bibinfo{person}{Abhishek
  Roy}, {and} \bibinfo{person}{Navrati Saxena}.}
  \bibinfo{year}{2016}\natexlab{}.
\newblock \showarticletitle{Next {{Generation 5G Wireless Networks}}: {{A
  Comprehensive Survey}}}.
\newblock \bibinfo{journal}{\emph{IEEE Communications Surveys \& Tutorials}}
  \bibinfo{volume}{18}, \bibinfo{number}{3} (\bibinfo{year}{2016}),
  \bibinfo{pages}{1617--1655}.
\newblock
\showISSN{1553-877X}


\bibitem[\protect\citeauthoryear{Andrews, Buzzi, Choi, Hanly, Lozano, Soong,
  and Zhang}{Andrews et~al\mbox{.}}{2014}]%
        {andrewsWhatWill5G2014}
\bibfield{author}{\bibinfo{person}{Jeffrey~G. Andrews},
  \bibinfo{person}{Stefano Buzzi}, \bibinfo{person}{Wan Choi},
  \bibinfo{person}{Stephen~V. Hanly}, \bibinfo{person}{Angel Lozano},
  \bibinfo{person}{Anthony C.~K. Soong}, {and}
  \bibinfo{person}{Jianzhong~Charlie Zhang}.} \bibinfo{year}{2014}\natexlab{}.
\newblock \showarticletitle{What {{Will 5G Be}}?}
\newblock \bibinfo{journal}{\emph{IEEE Journal on Selected Areas in
  Communications}} \bibinfo{volume}{32}, \bibinfo{number}{6}
  (\bibinfo{year}{2014}), \bibinfo{pages}{1065--1082}.
\newblock
\showISSN{0733-8716}


\bibitem[\protect\citeauthoryear{Arjoune and Faruque}{Arjoune and
  Faruque}{2020}]%
        {arjouneSmartJammingAttacks2020}
\bibfield{author}{\bibinfo{person}{Youness Arjoune} {and}
  \bibinfo{person}{Saleh Faruque}.} \bibinfo{year}{2020}\natexlab{}.
\newblock \showarticletitle{Smart {{Jamming Attacks}} in {{5G New Radio}}: {{A
  Review}}}. In \bibinfo{booktitle}{\emph{2020 10th {{Annual Computing}} and
  {{Communication Workshop}} and {{Conference}} ({{CCWC}})}} ({Las Vegas, NV,
  USA}). \bibinfo{publisher}{{IEEE}}, \bibinfo{pages}{1010--1015}.
\newblock
\showISBNx{978-1-72813-783-4}


\bibitem[\protect\citeauthoryear{Arun and Balakrishnan}{Arun and
  Balakrishnan}{2020}]%
        {arunRFocusBeamformingUsing2020}
\bibfield{author}{\bibinfo{person}{Venkat Arun} {and} \bibinfo{person}{Hari
  Balakrishnan}.} \bibinfo{year}{2020}\natexlab{}.
\newblock \showarticletitle{{{RFocus}}: {{Beamforming Using Thousands}} of
  {{Passive Antennas}}}. In \bibinfo{booktitle}{\emph{17th {{USENIX Symposium}}
  on {{Networked Systems Design}} and {{Implementation}} ({{NSDI}} 20)}}
  ({Santa Clara, CA}). \bibinfo{publisher}{{USENIX Association}},
  \bibinfo{pages}{1047--1061}.
\newblock
\showISBNx{978-1-939133-13-7}


\bibitem[\protect\citeauthoryear{Basar, Renzo, Rosny, Debbah, Alouini, and
  Zhang}{Basar et~al\mbox{.}}{2019}]%
        {basar_wireless_2019}
\bibfield{author}{\bibinfo{person}{E. Basar}, \bibinfo{person}{M.~Di Renzo},
  \bibinfo{person}{J.~De Rosny}, \bibinfo{person}{M. Debbah},
  \bibinfo{person}{M. Alouini}, {and} \bibinfo{person}{R. Zhang}.}
  \bibinfo{year}{2019}\natexlab{}.
\newblock \showarticletitle{Wireless {Communications} {Through}
  {Reconfigurable} {Intelligent} {Surfaces}}.
\newblock \bibinfo{journal}{\emph{IEEE Access}}  \bibinfo{volume}{7}
  (\bibinfo{year}{2019}), \bibinfo{pages}{116753--116773}.
\newblock


\bibitem[\protect\citeauthoryear{Björnson, Sanguinetti, Wymeersch, Hoydis, and
  Marzetta}{Björnson et~al\mbox{.}}{2019}]%
        {bjornsonMassiveMIMOReality2019}
\bibfield{author}{\bibinfo{person}{Emil Björnson}, \bibinfo{person}{Luca
  Sanguinetti}, \bibinfo{person}{Henk Wymeersch}, \bibinfo{person}{Jakob
  Hoydis}, {and} \bibinfo{person}{Thomas~L. Marzetta}.}
  \bibinfo{year}{2019}\natexlab{}.
\newblock \showarticletitle{Massive {{MIMO}} Is a Reality—{{What}} Is Next?}
\newblock \bibinfo{journal}{\emph{Digital Signal Processing}}
  \bibinfo{volume}{94} (\bibinfo{year}{2019}), \bibinfo{pages}{3--20}.
\newblock
\showISSN{10512004}


\bibitem[\protect\citeauthoryear{Chen, Liang, Pei, and Guo}{Chen
  et~al\mbox{.}}{2019}]%
        {chenIntelligentReflectingSurface2019}
\bibfield{author}{\bibinfo{person}{Jie Chen}, \bibinfo{person}{Ying-Chang
  Liang}, \bibinfo{person}{Yiyang Pei}, {and} \bibinfo{person}{Huayan Guo}.}
  \bibinfo{year}{2019}\natexlab{}.
\newblock \showarticletitle{Intelligent {{Reflecting Surface}}: {{A
  Programmable Wireless Environment}} for {{Physical Layer Security}}}.
\newblock \bibinfo{journal}{\emph{IEEE Access}}  \bibinfo{volume}{7}
  (\bibinfo{year}{2019}), \bibinfo{pages}{82599--82612}.
\newblock
\showISSN{2169-3536}


\bibitem[\protect\citeauthoryear{Chiang and Hu}{Chiang and Hu}{2011}]%
        {chiangCrossLayerJammingDetection2011}
\bibfield{author}{\bibinfo{person}{Jerry~T. Chiang} {and}
  \bibinfo{person}{Yih-Chun Hu}.} \bibinfo{year}{2011}\natexlab{}.
\newblock \showarticletitle{Cross-{{Layer Jamming Detection}} and
  {{Mitigation}} in {{Wireless Broadcast Networks}}}.
\newblock \bibinfo{journal}{\emph{IEEE/ACM Transactions on Networking}}
  \bibinfo{volume}{19}, \bibinfo{number}{1} (\bibinfo{year}{2011}),
  \bibinfo{pages}{286--298}.
\newblock
\showISSN{1063-6692, 1558-2566}


\bibitem[\protect\citeauthoryear{Chiueh, Tsai, {Lai. I-Wei}, and Chiueh}{Chiueh
  et~al\mbox{.}}{2012}]%
        {chiuehBasebandReceiverDesign2012}
\bibfield{author}{\bibinfo{person}{Tzi-Dar Chiueh}, \bibinfo{person}{Pei-Yun
  Tsai}, \bibinfo{person}{{Lai. I-Wei}}, {and} \bibinfo{person}{Tzi-Dar
  Chiueh}.} \bibinfo{year}{2012}\natexlab{}.
\newblock \bibinfo{booktitle}{\emph{Baseband Receiver Design for Wireless
  {{MIMO}}-{{OFDM}} Communications} (\bibinfo{edition}{2nd} ed.)}.
\newblock \bibinfo{publisher}{{Wiley}}, \bibinfo{address}{{Hoboken, N.J}}.
\newblock
\showISBNx{978-1-118-18821-7 978-1-118-18820-0}
\showLCCN{TK5103.2}


\bibitem[\protect\citeauthoryear{Clancy}{Clancy}{2011}]%
        {clancyEfficientOFDMDenial2011}
\bibfield{author}{\bibinfo{person}{T.~Charles Clancy}.}
  \bibinfo{year}{2011}\natexlab{}.
\newblock \showarticletitle{Efficient {{OFDM Denial}}: {{Pilot Jamming}} and
  {{Pilot Nulling}}}. In \bibinfo{booktitle}{\emph{2011 {{IEEE International
  Conference}} on {{Communications}} ({{ICC}})}} ({Kyoto, Japan}).
  \bibinfo{publisher}{{IEEE}}, \bibinfo{pages}{1--5}.
\newblock
\showISBNx{978-1-61284-232-5}


\bibitem[\protect\citeauthoryear{Coleri, Ergen, Puri, and Bahai}{Coleri
  et~al\mbox{.}}{2002}]%
        {coleriChannelEstimationTechniques2002}
\bibfield{author}{\bibinfo{person}{S. Coleri}, \bibinfo{person}{M. Ergen},
  \bibinfo{person}{A. Puri}, {and} \bibinfo{person}{A. Bahai}.}
  \bibinfo{year}{2002}\natexlab{}.
\newblock \showarticletitle{{Channel Estimation Techniques Based on Pilot
  Arrangement in {{OFDM}} Systems}}.
\newblock \bibinfo{journal}{\emph{IEEE Transactions on Broadcasting}}
  \bibinfo{volume}{48}, \bibinfo{number}{3} (\bibinfo{date}{Sept.}
  \bibinfo{year}{2002}), \bibinfo{pages}{223--229}.
\newblock
\showISSN{0018-9316}


\bibitem[\protect\citeauthoryear{Cui, Zhang, and Zhang}{Cui
  et~al\mbox{.}}{2019}]%
        {cuiSecureWirelessCommunication2019}
\bibfield{author}{\bibinfo{person}{Miao Cui}, \bibinfo{person}{Guangchi Zhang},
  {and} \bibinfo{person}{Rui Zhang}.} \bibinfo{year}{2019}\natexlab{}.
\newblock \showarticletitle{Secure {{Wireless Communication}} via {{Intelligent
  Reflecting Surface}}}.
\newblock \bibinfo{journal}{\emph{IEEE Wireless Communications Letters}}
  \bibinfo{volume}{8}, \bibinfo{number}{5} (\bibinfo{year}{2019}),
  \bibinfo{pages}{1410--1414}.
\newblock
\showISSN{2162-2337, 2162-2345}


\bibitem[\protect\citeauthoryear{del Hougne, Fink, and Lerosey}{del Hougne
  et~al\mbox{.}}{2019}]%
        {del_hougne_optimally_2019}
\bibfield{author}{\bibinfo{person}{Philipp del Hougne},
  \bibinfo{person}{Mathias Fink}, {and} \bibinfo{person}{Geoffroy Lerosey}.}
  \bibinfo{year}{2019}\natexlab{}.
\newblock \showarticletitle{Optimally Diverse Communication Channels in
  Disordered Environments with Tuned Randomness}.
\newblock \bibinfo{journal}{\emph{Nature Electronics}} \bibinfo{volume}{2},
  \bibinfo{number}{1} (\bibinfo{year}{2019}), \bibinfo{pages}{36--41}.
\newblock
\showISSN{2520-1131}


\bibitem[\protect\citeauthoryear{ETSI}{ETSI}{2018}]%
        {etsiETSITS1382018}
\bibfield{author}{\bibinfo{person}{ETSI}.} \bibinfo{year}{2018}\natexlab{}.
\newblock \bibinfo{title}{{{ETSI TS}} 138 214 {{V15}}.2.0, {{5G}}; {{NR}};
  {{Physical}} Layer Procedures for Data}.
\newblock
\newblock


\bibitem[\protect\citeauthoryear{Girke, Kurtz, Dorsch, and Wietfeld}{Girke
  et~al\mbox{.}}{2019}]%
        {girkeResilient5GLessons2019}
\bibfield{author}{\bibinfo{person}{Felix Girke}, \bibinfo{person}{Fabian
  Kurtz}, \bibinfo{person}{Nils Dorsch}, {and} \bibinfo{person}{Christian
  Wietfeld}.} \bibinfo{year}{2019}\natexlab{}.
\newblock \showarticletitle{Towards {{Resilient 5G}}: {{Lessons Learned}} from
  {{Experimental Evaluations}} of {{LTE Uplink Jamming}}}. In
  \bibinfo{booktitle}{\emph{2019 {{IEEE International Conference}} on
  {{Communications Workshops}} ({{ICC Workshops}})}} ({Shanghai, China}).
  \bibinfo{publisher}{{IEEE}}, \bibinfo{pages}{1--6}.
\newblock
\showISBNx{978-1-72812-373-8}


\bibitem[\protect\citeauthoryear{Goldsmith}{Goldsmith}{2005}]%
        {goldsmith_wireless_2005}
\bibfield{author}{\bibinfo{person}{Andrea Goldsmith}.}
  \bibinfo{year}{2005}\natexlab{}.
\newblock \bibinfo{booktitle}{\emph{Wireless {Communications}}}.
\newblock \bibinfo{publisher}{Cambridge University Press},
  \bibinfo{address}{USA}.
\newblock
\showISBNx{0-521-83716-2}


\bibitem[\protect\citeauthoryear{Greenerwave}{Greenerwave}{[n.d.]}]%
        {greenerwave_website}
\bibfield{author}{\bibinfo{person}{Greenerwave}.}
  \bibinfo{year}{[n.d.]}\natexlab{}.
\newblock
\newblock
\urldef\tempurl%
\url{http://greenerwave.com/}
\showURL{%
\tempurl}
\newblock
\shownote{{Accessed: July 30, 2021}.}


\bibitem[\protect\citeauthoryear{Grover, Lim, and Yang}{Grover
  et~al\mbox{.}}{2014}]%
        {groverJammingAntijammingTechniques2014}
\bibfield{author}{\bibinfo{person}{Kanika Grover}, \bibinfo{person}{Alvin Lim},
  {and} \bibinfo{person}{Qing Yang}.} \bibinfo{year}{2014}\natexlab{}.
\newblock \showarticletitle{Jamming and Anti-Jamming Techniques in Wireless
  Networks: A Survey}.
\newblock \bibinfo{journal}{\emph{International Journal of Ad Hoc and
  Ubiquitous Computing}} \bibinfo{volume}{17}, \bibinfo{number}{4}
  (\bibinfo{year}{2014}), \bibinfo{pages}{197}.
\newblock
\showISSN{1743-8225, 1743-8233}


\bibitem[\protect\citeauthoryear{Gupta and Jha}{Gupta and Jha}{2015}]%
        {guptaSurvey5GNetwork2015}
\bibfield{author}{\bibinfo{person}{A. Gupta} {and} \bibinfo{person}{R.~K.
  Jha}.} \bibinfo{year}{2015}\natexlab{}.
\newblock \showarticletitle{A {{Survey}} of {{5G Network}}: {{Architecture}}
  and {{Emerging Technologies}}}.
\newblock \bibinfo{journal}{\emph{IEEE Access}}  \bibinfo{volume}{3}
  (\bibinfo{year}{2015}), \bibinfo{pages}{1206--1232}.
\newblock
\showISSN{2169-3536}


\bibitem[\protect\citeauthoryear{Hang, Zanji, and Jingbo}{Hang
  et~al\mbox{.}}{2006}]%
        {hangPerformanceDSSSRepeater2006}
\bibfield{author}{\bibinfo{person}{Wang Hang}, \bibinfo{person}{Wang Zanji},
  {and} \bibinfo{person}{Guo Jingbo}.} \bibinfo{year}{2006}\natexlab{}.
\newblock \showarticletitle{Performance of {{DSSS}} against {{Repeater
  Jamming}}}. In \bibinfo{booktitle}{\emph{2006 13th {{IEEE International
  Conference}} on {{Electronics}}, {{Circuits}} and {{Systems}}}}.
  \bibinfo{publisher}{{IEEE}}, \bibinfo{address}{{Nice, France}},
  \bibinfo{pages}{858--861}.
\newblock
\showISBNx{978-1-4244-0394-3 978-1-4244-0395-0}


\bibitem[\protect\citeauthoryear{Huang and Wang}{Huang and Wang}{2021}]%
        {huangIntelligentReflectingSurface2020}
\bibfield{author}{\bibinfo{person}{Ke-Wen Huang} {and}
  \bibinfo{person}{Hui-Ming Wang}.} \bibinfo{year}{2021}\natexlab{}.
\newblock \showarticletitle{Intelligent {{Reflecting Surface Aided Pilot
  Contamination Attack}} and {{Its Countermeasure}}}.
\newblock \bibinfo{journal}{\emph{IEEE Transactions on Wireless
  Communications}} \bibinfo{volume}{20}, \bibinfo{number}{1}
  (\bibinfo{year}{2021}), \bibinfo{pages}{345--359}.
\newblock


\bibitem[\protect\citeauthoryear{Hum and Perruisseau-Carrier}{Hum and
  Perruisseau-Carrier}{2014}]%
        {humReconfigurableReflectarraysArray2014}
\bibfield{author}{\bibinfo{person}{Sean~Victor Hum} {and}
  \bibinfo{person}{Julien Perruisseau-Carrier}.}
  \bibinfo{year}{2014}\natexlab{}.
\newblock \showarticletitle{Reconfigurable {{Reflectarrays}} and {{Array
  Lenses}} for {{Dynamic Antenna Beam Control}}: {{A Review}}}.
\newblock \bibinfo{journal}{\emph{IEEE Transactions on Antennas and
  Propagation}} \bibinfo{volume}{62}, \bibinfo{number}{1}
  (\bibinfo{year}{2014}), \bibinfo{pages}{183--198}.
\newblock
\showISSN{0018-926X, 1558-2221}


\bibitem[\protect\citeauthoryear{IEEE}{IEEE}{2013}]%
        {IEEEStandardInformation}
\bibfield{author}{\bibinfo{person}{IEEE}.} \bibinfo{year}{2013}\natexlab{}.
\newblock \bibinfo{title}{{Telecommunications and information exchange between
  systems Local and metropolitan area networks-- Specific requirements Part 11:
  Wireless LAN Medium Access Control (MAC) and Physical Layer (PHY)
  Specifications Amendment 4: Enhancements for Very HighThroughput for
  Operation in Bands below 6 GHz}}.
\newblock
\newblock
\showISBNx{9780738188607}
\newblock
\shownote{{Accessed: July 30, 2021}.}


\bibitem[\protect\citeauthoryear{Kaina, Dupré, Lerosey, and Fink}{Kaina
  et~al\mbox{.}}{2015}]%
        {kaina_shaping_2015}
\bibfield{author}{\bibinfo{person}{Nadège Kaina}, \bibinfo{person}{Matthieu
  Dupré}, \bibinfo{person}{Geoffroy Lerosey}, {and} \bibinfo{person}{Mathias
  Fink}.} \bibinfo{year}{2015}\natexlab{}.
\newblock \showarticletitle{Shaping complex microwave fields in reverberating
  media with binary tunable metasurfaces}.
\newblock \bibinfo{journal}{\emph{Scientific Reports}} \bibinfo{volume}{4},
  \bibinfo{number}{1} (\bibinfo{date}{May} \bibinfo{year}{2015}),
  \bibinfo{pages}{6693}.
\newblock
\showISSN{2045-2322}


\bibitem[\protect\citeauthoryear{Liaskos et~al\mbox{.}}{Liaskos
  et~al\mbox{.}}{2019}]%
        {liaskos_novel_2019}
\bibfield{author}{\bibinfo{person}{Christos Liaskos} {et~al\mbox{.}}}
  \bibinfo{year}{2019}\natexlab{}.
\newblock \showarticletitle{A novel communication paradigm for high capacity
  and security via programmable indoor wireless environments in next generation
  wireless systems}.
\newblock \bibinfo{journal}{\emph{Ad Hoc Networks}}  \bibinfo{volume}{87}
  (\bibinfo{date}{May} \bibinfo{year}{2019}), \bibinfo{pages}{1--16}.
\newblock
\showISSN{15708705}


\bibitem[\protect\citeauthoryear{Lichtman, Jover, Labib, Rao, Marojevic, and
  Reed}{Lichtman et~al\mbox{.}}{2016}]%
        {lichtmanLTELTEAJamming2016}
\bibfield{author}{\bibinfo{person}{Marc Lichtman},
  \bibinfo{person}{Roger~Piqueras Jover}, \bibinfo{person}{Mina Labib},
  \bibinfo{person}{Raghunandan Rao}, \bibinfo{person}{Vuk Marojevic}, {and}
  \bibinfo{person}{Jeffrey~H. Reed}.} \bibinfo{year}{2016}\natexlab{}.
\newblock \showarticletitle{{{LTE}}/{{LTE}}-{{A}} Jamming, Spoofing, and
  Sniffing: Threat Assessment and Mitigation}.
\newblock \bibinfo{journal}{\emph{IEEE Communications Magazine}}
  \bibinfo{volume}{54}, \bibinfo{number}{4} (\bibinfo{year}{2016}),
  \bibinfo{pages}{54--61}.
\newblock
\showISSN{0163-6804}


\bibitem[\protect\citeauthoryear{Lichtman, Poston, Amuru, Shahriar, Clancy,
  Buehrer, and Reed}{Lichtman et~al\mbox{.}}{6 01}]%
        {lichtmanCommunicationsJammingTaxonomy2016}
\bibfield{author}{\bibinfo{person}{Marc Lichtman}, \bibinfo{person}{Jeffrey~D.
  Poston}, \bibinfo{person}{SaiDhiraj Amuru}, \bibinfo{person}{Chowdhury
  Shahriar}, \bibinfo{person}{T.~Charles Clancy}, \bibinfo{person}{R.~Michael
  Buehrer}, {and} \bibinfo{person}{Jeffrey~H. Reed}.}
  \bibinfo{year}{2016-01}\natexlab{}.
\newblock \showarticletitle{A Communications Jamming Taxonomy}.
\newblock \bibinfo{journal}{\emph{IEEE Security \& Privacy}}
  \bibinfo{volume}{14}, \bibinfo{number}{1} (\bibinfo{year}{2016-01}),
  \bibinfo{pages}{47--54}.
\newblock
\showISSN{1540-7993}


\bibitem[\protect\citeauthoryear{Lyamin, Vinel, Jonsson, and Loo}{Lyamin
  et~al\mbox{.}}{2014}]%
        {lyaminRealTimeDetectionDenialofService2014}
\bibfield{author}{\bibinfo{person}{Nikita Lyamin}, \bibinfo{person}{Alexey
  Vinel}, \bibinfo{person}{Magnus Jonsson}, {and} \bibinfo{person}{Jonathan
  Loo}.} \bibinfo{year}{2014}\natexlab{}.
\newblock \showarticletitle{Real-{{Time Detection}} of {{Denial}}-of-{{Service
  Attacks}} in {{IEEE}} 802.11p {{Vehicular Networks}}}.
\newblock \bibinfo{journal}{\emph{IEEE Communications Letters}}
  \bibinfo{volume}{18}, \bibinfo{number}{1} (\bibinfo{year}{2014}),
  \bibinfo{pages}{110--113}.
\newblock
\showISSN{1089-7798}


\bibitem[\protect\citeauthoryear{Lyu, Hoang, Gong, Niyato, and Kim}{Lyu
  et~al\mbox{.}}{2020}]%
        {lyuIRSBasedWirelessJamming2020}
\bibfield{author}{\bibinfo{person}{Bin Lyu}, \bibinfo{person}{Dinh~Thai Hoang},
  \bibinfo{person}{Shimin Gong}, \bibinfo{person}{Dusit Niyato}, {and}
  \bibinfo{person}{Dong~In Kim}.} \bibinfo{year}{2020}\natexlab{}.
\newblock \showarticletitle{{{IRS}}-{{Based Wireless Jamming Attacks}}: {{When
  Jammers Can Attack Without Power}}}.
\newblock \bibinfo{journal}{\emph{IEEE Wireless Communications Letters}}
  \bibinfo{volume}{9}, \bibinfo{number}{10} (\bibinfo{date}{Oct.}
  \bibinfo{year}{2020}), \bibinfo{pages}{1663--1667}.
\newblock
\showISSN{2162-2337, 2162-2345}


\bibitem[\protect\citeauthoryear{MathWorks}{MathWorks}{[n.d.]}]%
        {wlan_tb_website}
\bibfield{author}{\bibinfo{person}{MathWorks}.}
  \bibinfo{year}{[n.d.]}\natexlab{}.
\newblock \bibinfo{title}{{WLAN Toolbox - MATLAB}}.
\newblock
\newblock
\urldef\tempurl%
\url{https://www.mathworks.com/products/wlan.html}
\showURL{%
\tempurl}
\newblock
\shownote{{Accessed: July 30, 2021}.}


\bibitem[\protect\citeauthoryear{{Metawave Corporation}}{{Metawave
  Corporation}}{[n.d.]}]%
        {metawave_website}
\bibfield{author}{\bibinfo{person}{{Metawave Corporation}}.}
  \bibinfo{year}{[n.d.]}\natexlab{}.
\newblock
\newblock
\urldef\tempurl%
\url{https://www.metawave.co/}
\showURL{%
\tempurl}
\newblock
\shownote{{Accessed: July 30, 2021}.}


\bibitem[\protect\citeauthoryear{{\"O}zdogan, Bj{\"o}rnson, and
  Larsson}{{\"O}zdogan et~al\mbox{.}}{2020}]%
        {ozdoganIntelligentReflectingSurfaces2020}
\bibfield{author}{\bibinfo{person}{{\"O}zgecan {\"O}zdogan},
  \bibinfo{person}{Emil Bj{\"o}rnson}, {and} \bibinfo{person}{Erik~G.
  Larsson}.} \bibinfo{year}{2020}\natexlab{}.
\newblock \showarticletitle{Intelligent {{Reflecting Surfaces}}: {{Physics}},
  {{Propagation}}, and {{Pathloss Modeling}}}.
\newblock \bibinfo{journal}{\emph{IEEE Wireless Communications Letters}}
  \bibinfo{volume}{9}, \bibinfo{number}{5} (\bibinfo{date}{May}
  \bibinfo{year}{2020}), \bibinfo{pages}{581--585}.
\newblock
\showISSN{2162-2337, 2162-2345}


\bibitem[\protect\citeauthoryear{Pei, Yin, Tan, Cao, Li, Wang, Zhang, and
  Björnson}{Pei et~al\mbox{.}}{2021}]%
        {peiRISAidedWirelessCommunications2021}
\bibfield{author}{\bibinfo{person}{Xilong Pei}, \bibinfo{person}{Haifan Yin},
  \bibinfo{person}{Li Tan}, \bibinfo{person}{Lin Cao},
  \bibinfo{person}{Zhanpeng Li}, \bibinfo{person}{Kai Wang},
  \bibinfo{person}{Kun Zhang}, {and} \bibinfo{person}{Emil Björnson}.}
  \bibinfo{year}{2021}\natexlab{}.
\newblock \showarticletitle{{{RIS}}-{{Aided Wireless Communications}}:
  {{Prototyping}}, {{Adaptive Beamforming}}, and {{Indoor}}/{{Outdoor Field
  Trials}}}.
\newblock  (\bibinfo{year}{2021}).
\newblock
\showeprint{2103.00534}


\bibitem[\protect\citeauthoryear{Poisel}{Poisel}{2011}]%
        {poiselModernCommunicationsJamming2011}
\bibfield{author}{\bibinfo{person}{Richard Poisel}.}
  \bibinfo{year}{2011}\natexlab{}.
\newblock \bibinfo{booktitle}{\emph{Modern Communications Jamming: Principles
  and Techniques} (\bibinfo{edition}{2nd} ed.)}.
\newblock \bibinfo{publisher}{{Artech House}}.
\newblock
\showISBNx{978-1-60807-165-4}


\bibitem[\protect\citeauthoryear{P{\"o}pper, Strasser, and {\v
  C}apkun}{P{\"o}pper et~al\mbox{.}}{2009}]%
        {popperJammingResistantBroadcastCommunication2009}
\bibfield{author}{\bibinfo{person}{Christina P{\"o}pper},
  \bibinfo{person}{Mario Strasser}, {and} \bibinfo{person}{Srdjan {\v
  C}apkun}.} \bibinfo{year}{2009}\natexlab{}.
\newblock \showarticletitle{Jamming-{{Resistant Broadcast Communication}}
  without {{Shared Keys}}}. In \bibinfo{booktitle}{\emph{Proceedings of the
  18th {{Conference}} on {{USENIX Security Symposium}}}}.
  \bibinfo{publisher}{{USENIX Association}}, \bibinfo{address}{{USA}},
  \bibinfo{pages}{231--248}.
\newblock


\bibitem[\protect\citeauthoryear{Renzo, Debbah, Phan-Huy, Zappone, Alouini,
  Yuen, Sciancalepore, Alexandropoulos, Hoydis, Gacanin, de~Rosny, Bounceur,
  Lerosey, and Fink}{Renzo et~al\mbox{.}}{2019}]%
        {renzoSmartRadioEnvironments2019}
\bibfield{author}{\bibinfo{person}{Marco~Di Renzo}, \bibinfo{person}{Merouane
  Debbah}, \bibinfo{person}{Dinh-Thuy Phan-Huy}, \bibinfo{person}{Alessio
  Zappone}, \bibinfo{person}{Mohamed-Slim Alouini}, \bibinfo{person}{Chau
  Yuen}, \bibinfo{person}{Vincenzo Sciancalepore}, \bibinfo{person}{George~C.
  Alexandropoulos}, \bibinfo{person}{Jakob Hoydis}, \bibinfo{person}{Haris
  Gacanin}, \bibinfo{person}{Julien de Rosny}, \bibinfo{person}{Ahcene
  Bounceur}, \bibinfo{person}{Geoffroy Lerosey}, {and} \bibinfo{person}{Mathias
  Fink}.} \bibinfo{year}{2019}\natexlab{}.
\newblock \showarticletitle{Smart Radio Environments Empowered by
  Reconfigurable {{AI}} Meta-Surfaces: An Idea Whose Time Has Come}.
\newblock \bibinfo{journal}{\emph{EURASIP Journal on Wireless Communications
  and Networking}} \bibinfo{volume}{2019}, \bibinfo{number}{1}
  (\bibinfo{year}{2019}), \bibinfo{pages}{129}.
\newblock
\showISSN{1687-1499}


\bibitem[\protect\citeauthoryear{Roth, Tomasin, Maso, and Sezgin}{Roth
  et~al\mbox{.}}{2021}]%
        {rothLocalizationAttackPrecoder2020}
\bibfield{author}{\bibinfo{person}{Stefan Roth}, \bibinfo{person}{Stefano
  Tomasin}, \bibinfo{person}{Marco Maso}, {and} \bibinfo{person}{Aydin
  Sezgin}.} \bibinfo{year}{2021}\natexlab{}.
\newblock \showarticletitle{Localization Attack by Precoder Feedback
  Overhearing in 5G Networks and Countermeasures}.
\newblock \bibinfo{journal}{\emph{IEEE Transactions on Wireless
  Communications}} \bibinfo{volume}{20}, \bibinfo{number}{7}
  (\bibinfo{year}{2021}), \bibinfo{pages}{4100--4112}.
\newblock


\bibitem[\protect\citeauthoryear{{speedtest-cli}}{{speedtest-cli}}{[n.d.]}]%
        {speedtest_cli_git}
\bibfield{author}{\bibinfo{person}{{speedtest-cli}}.}
  \bibinfo{year}{[n.d.]}\natexlab{}.
\newblock
\newblock
\urldef\tempurl%
\url{https://github.com/sivel/speedtest-cli}
\showURL{%
\tempurl}
\newblock
\shownote{{Accessed: July 30, 2021}.}


\bibitem[\protect\citeauthoryear{Strasser, Danev, and {\v C}apkun}{Strasser
  et~al\mbox{.}}{2010}]%
        {strasserDetectionReactiveJamming2010}
\bibfield{author}{\bibinfo{person}{Mario Strasser}, \bibinfo{person}{Boris
  Danev}, {and} \bibinfo{person}{Srdjan {\v C}apkun}.}
  \bibinfo{year}{2010}\natexlab{}.
\newblock \showarticletitle{Detection of Reactive Jamming in Sensor Networks}.
\newblock \bibinfo{journal}{\emph{ACM Transactions on Sensor Networks}}
  \bibinfo{volume}{7}, \bibinfo{number}{2} (\bibinfo{year}{2010}),
  \bibinfo{pages}{1--29}.
\newblock
\showISSN{1550-4859, 1550-4867}


\bibitem[\protect\citeauthoryear{Tippenhauer, Malisa, Ranganathan, and
  Capkun}{Tippenhauer et~al\mbox{.}}{2013}]%
        {tippenhauerLimitationsFriendlyJamming2013}
\bibfield{author}{\bibinfo{person}{N.~O. Tippenhauer}, \bibinfo{person}{L.
  Malisa}, \bibinfo{person}{A. Ranganathan}, {and} \bibinfo{person}{S.
  Capkun}.} \bibinfo{year}{2013}\natexlab{}.
\newblock \showarticletitle{On {{Limitations}} of {{Friendly Jamming}} for
  {{Confidentiality}}}. In \bibinfo{booktitle}{\emph{2013 {{IEEE Symposium}} on
  {{Security}} and {{Privacy}}}} ({Berkeley, CA}). \bibinfo{publisher}{{IEEE}},
  \bibinfo{pages}{160--173}.
\newblock
\showISBNx{978-0-7695-4977-4 978-1-4673-6166-8}


\bibitem[\protect\citeauthoryear{Vergès}{Vergès}{[n.d.]}]%
        {mcs_index_website}
\bibfield{author}{\bibinfo{person}{François Vergès}.}
  \bibinfo{year}{[n.d.]}\natexlab{}.
\newblock \bibinfo{title}{{MCS Index, Modulation and Coding Index 11n and
  11ac}}.
\newblock
\newblock
\urldef\tempurl%
\url{http://mcsindex.com/}
\showURL{%
\tempurl}
\newblock
\shownote{{Accessed: July 30, 2021}.}


\bibitem[\protect\citeauthoryear{{Wenbo Shen}, {Peng Ning}, {Xiaofan He}, and
  {Huaiyu Dai}}{{Wenbo Shen} et~al\mbox{.}}{2013}]%
        {wenboshenAllyFriendlyJamming2013}
\bibfield{author}{\bibinfo{person}{{Wenbo Shen}}, \bibinfo{person}{{Peng
  Ning}}, \bibinfo{person}{{Xiaofan He}}, {and} \bibinfo{person}{{Huaiyu
  Dai}}.} \bibinfo{year}{2013}\natexlab{}.
\newblock \showarticletitle{Ally {{Friendly Jamming}}: {{How}} to {{Jam Your
  Enemy}} and {{Maintain Your Own Wireless Connectivity}} at the {{Same
  Time}}}. In \bibinfo{booktitle}{\emph{2013 {{IEEE Symposium}} on {{Security}}
  and {{Privacy}}}} ({Berkeley, CA}). \bibinfo{publisher}{{IEEE}},
  \bibinfo{pages}{174--188}.
\newblock
\showISBNx{978-0-7695-4977-4 978-1-4673-6166-8}


\bibitem[\protect\citeauthoryear{Wu and Zhang}{Wu and Zhang}{2020}]%
        {wuSmartReconfigurableEnvironment2020}
\bibfield{author}{\bibinfo{person}{Qingqing Wu} {and} \bibinfo{person}{Rui
  Zhang}.} \bibinfo{year}{2020}\natexlab{}.
\newblock \showarticletitle{Towards {{Smart}} and {{Reconfigurable
  Environment}}: {{Intelligent Reflecting Surface Aided Wireless Network}}}.
\newblock \bibinfo{journal}{\emph{IEEE Communications Magazine}}
  \bibinfo{volume}{58}, \bibinfo{number}{1} (\bibinfo{year}{2020}),
  \bibinfo{pages}{106--112}.
\newblock
\showISSN{0163-6804, 1558-1896}


\bibitem[\protect\citeauthoryear{Wu, Zhang, Zheng, You, and Zhang}{Wu
  et~al\mbox{.}}{2021}]%
        {wuIntelligentReflectingSurface2021}
\bibfield{author}{\bibinfo{person}{Qingqing Wu}, \bibinfo{person}{Shuowen
  Zhang}, \bibinfo{person}{Beixiong Zheng}, \bibinfo{person}{Changsheng You},
  {and} \bibinfo{person}{Rui Zhang}.} \bibinfo{year}{2021}\natexlab{}.
\newblock \showarticletitle{Intelligent {{Reflecting Surface Aided Wireless
  Communications}}: {{A Tutorial}}}.
\newblock \bibinfo{journal}{\emph{IEEE Transactions on Communications}}
  \bibinfo{volume}{69}, \bibinfo{number}{5} (\bibinfo{year}{2021}),
  \bibinfo{pages}{3313--3351}.
\newblock
\showISSN{0090-6778, 1558-0857}


\bibitem[\protect\citeauthoryear{Xie, Li, and Li}{Xie et~al\mbox{.}}{2015}]%
        {xie_precise_2015}
\bibfield{author}{\bibinfo{person}{Yaxiong Xie}, \bibinfo{person}{Zhenjiang
  Li}, {and} \bibinfo{person}{Mo Li}.} \bibinfo{year}{2015}\natexlab{}.
\newblock \showarticletitle{Precise {Power} {Delay} {Profiling} with
  {Commodity} {WiFi}}. In \bibinfo{booktitle}{\emph{Proceedings of the 21st
  {Annual} {International} {Conference} on {Mobile} {Computing} and
  {Networking}}} \emph{(\bibinfo{series}{{MobiCom} '15})}.
  \bibinfo{publisher}{ACM}, \bibinfo{address}{New York, NY, USA},
  \bibinfo{pages}{53--64}.
\newblock
\showISBNx{978-1-4503-3619-2}
\newblock
\shownote{{Paris, France}.}


\bibitem[\protect\citeauthoryear{Yang, Cao, Yang, Gao, Xu, Li, Chen, Zhao,
  Zheng, and Li}{Yang et~al\mbox{.}}{6 12}]%
        {yangProgrammableMetasurfaceDynamic2016}
\bibfield{author}{\bibinfo{person}{Huanhuan Yang}, \bibinfo{person}{Xiangyu
  Cao}, \bibinfo{person}{Fan Yang}, \bibinfo{person}{Jun Gao},
  \bibinfo{person}{Shenheng Xu}, \bibinfo{person}{Maokun Li},
  \bibinfo{person}{Xibi Chen}, \bibinfo{person}{Yi Zhao},
  \bibinfo{person}{Yuejun Zheng}, {and} \bibinfo{person}{Sijia Li}.}
  \bibinfo{year}{2016-12}\natexlab{}.
\newblock \showarticletitle{{A Programmable Metasurface with Dynamic
  Polarization, Scattering and Focusing Control}}.
\newblock \bibinfo{journal}{\emph{Scientific Reports}} \bibinfo{volume}{6},
  \bibinfo{number}{1} (\bibinfo{year}{2016-12}), \bibinfo{pages}{35692}.
\newblock
\showISSN{2045-2322}


\bibitem[\protect\citeauthoryear{Yang, Xiong, Zhao, Niyato, Wu, Poor, and
  Tornatore}{Yang et~al\mbox{.}}{2021}]%
        {yangIntelligentReflectingSurface2021}
\bibfield{author}{\bibinfo{person}{Helin Yang}, \bibinfo{person}{Zehui Xiong},
  \bibinfo{person}{Jun Zhao}, \bibinfo{person}{Dusit Niyato},
  \bibinfo{person}{Qingqing Wu}, \bibinfo{person}{H.~Vincent Poor}, {and}
  \bibinfo{person}{Massimo Tornatore}.} \bibinfo{year}{2021}\natexlab{}.
\newblock \showarticletitle{Intelligent {{Reflecting Surface Assisted
  Anti}}-{{Jamming Communications}}: {{A Fast Reinforcement Learning
  Approach}}}.
\newblock \bibinfo{journal}{\emph{IEEE Transactions on Wireless
  Communications}} \bibinfo{volume}{20}, \bibinfo{number}{3}
  (\bibinfo{year}{2021}), \bibinfo{pages}{1963--1974}.
\newblock
\showISSN{1536-1276, 1558-2248}


\bibitem[\protect\citeauthoryear{Yang, Yang, Xu, Mao, Li, Cao, and Gao}{Yang
  et~al\mbox{.}}{2016}]%
        {yang1Bit10Times2016}
\bibfield{author}{\bibinfo{person}{Huanhuan Yang}, \bibinfo{person}{Fan Yang},
  \bibinfo{person}{Shenheng Xu}, \bibinfo{person}{Yilin Mao},
  \bibinfo{person}{Maokun Li}, \bibinfo{person}{Xiangyu Cao}, {and}
  \bibinfo{person}{Jun Gao}.} \bibinfo{year}{2016}\natexlab{}.
\newblock \showarticletitle{A 1-{{Bit}} $10 \times 10$ {{Reconfigurable
  Reflectarray Antenna}}: {{Design}}, {{Optimization}}, and {{Experiment}}}.
\newblock \bibinfo{journal}{\emph{IEEE Transactions on Antennas and
  Propagation}} \bibinfo{volume}{64}, \bibinfo{number}{6}
  (\bibinfo{year}{2016}), \bibinfo{pages}{2246--2254}.
\newblock
\showISSN{0018-926X, 1558-2221}


\bibitem[\protect\citeauthoryear{Yang, Xiao, Xiao, and Li}{Yang
  et~al\mbox{.}}{2019}]%
        {yang6GWirelessCommunications2019}
\bibfield{author}{\bibinfo{person}{Ping Yang}, \bibinfo{person}{Yue Xiao},
  \bibinfo{person}{Ming Xiao}, {and} \bibinfo{person}{Shaoqian Li}.}
  \bibinfo{year}{2019}\natexlab{}.
\newblock \showarticletitle{{{6G Wireless Communications}}: {{Vision}} and
  {{Potential Techniques}}}.
\newblock \bibinfo{journal}{\emph{IEEE Network}} \bibinfo{volume}{33},
  \bibinfo{number}{4} (\bibinfo{date}{July} \bibinfo{year}{2019}),
  \bibinfo{pages}{70--75}.
\newblock
\showISSN{0890-8044, 1558-156X}


\end{thebibliography}

\appendix

\section{Derivation of ICI Power}
\label{appendix:ici}
We here derive the ICI arising from the ERA due to sub-symbol channel variations. Fortunately, $H_{k,k'}[n]$ can be related to the complex time varying channel impulse response~(CIR) $h_l[n, m]$, at the $m^{th}$ sample of the $n^{th}$ OFDM-symbol for all $L$, $l= 0,\ldots, L-1$, channel taps~\cite{chiuehBasebandReceiverDesign2012}:
\begin{equation}
    H_{k,k'}[n] = \frac{1}{K} \sum_{l=0}^{L-1}   \underbrace{ \sum_{m=0}^{K-1} h_l[n,m]\ e^{-j 2 \pi m (k- k')/K}}_{H_l[n,k- k']}  \cdot e^{-j2\pi l k' / K}
    \label{eq:ici_vs_dk}
\end{equation}
where $H_l[n,k- k']$ is the discrete Fourier transform (DFT) of the $l^{th}$ channel tap in time (sample) direction at the subcarrier offset~$k - k'$. While static channels do not result in any ICI, the frequency contents of the fluctuating channel response during the OFDM symbol yield crosstalk from offset subcarriers $k'$. Note that for the desired signal, \ie, $k' = k$, (\ref{eq:ici_vs_dk}) yields the channel frequency response of the time-averaged CIR. During the ERA, the attacker switches between IRS surface configurations. Naturally, switching corresponds to abrupt changes within the channel response of Alice and Bob, and therefore we expect  $H_l[n,k- k']$ to contain significant high-frequency terms. We now will continue showing that the ERA is capable of turning the complete signal power from the attacker to interference. 
We account for the attacker's IRS by splitting the CIR into static direct (non-IRS) and IRS portions:
\begin{equation}
    h_l[n,m] = h^{d}_{l} + h^{IRS}_{l}[n,m].
    \label{eq:cir_with_IRS}
\end{equation}
Assuming that the attacker only affects a single channel tap $l = l_{IRS}$, the IRS-induced ICI is thus found from (\ref{eq:ici_vs_dk}), omitting the non-IRS taps: 
\begin{align}
    H^{IRS}_{k,k'}[n] & = \frac{1}{K}  H_{l_{IRS}}[n,k- k'] \cdot e^{-j2\pi l_{IRS} k' / K}\label{eq:ici_sp},
\end{align}
with squared magnitude given by
\begin{equation}
    \left|H^{IRS}_{k,k'}[n] \right|^2=  \frac{1}{K^2} \left| H_{l_{IRS}}[n,k- k'] \right|^2\label{eq:ici_mag2}.
\end{equation}
For brevity and simplicity, we here consider the special case that the IRS is configured such that the sum of the IRS channel tap over one OFDM symbol is zero, namely 
\begin{equation}
    \sum_{m=0}^{K-1} h_{l_{IRS}}[n,m] =  H_{l_{IRS}}[n,0] =0 \label{eq:ici_condition1}.
\end{equation}
Substituting this in (\ref{eq:ici_sp}) and setting $k’=k$ results in
\begin{align}
    H^{IRS}_{k}[n] &= H^{IRS}_{k,k}[n] = \frac{1}{K}  H_{l_{IRS}}[n,0] \cdot e^{-j2\pi l_{IRS} k / K}=0\label{eq:ici_sp_0},
\end{align}
which means that the IRS channel tap does not contribute to the useful signal but to the ICI only. 
Using (\ref{eq:h_eff}), the signal power of the useful signal $S_k$ is thus given by:
\begin{equation}
S_k = \left|H_k[n]\right|^2 = \left|H_k^{IRS}[n] + d_k \right|^2 = \left|d_k\right|^2.
\end{equation}
Assuming that all data symbols $X_{k}[n]$ on different subcarriers and OFDM symbols are independent and using (\ref{eq:ici_mag2}) and (\ref{eq:ici_sp_0}), the total ICI power due to the IRS is given by
\begin{align}
    I_{IRS}&= \sum_{k' \neq k}\left|H^{IRS}_{k,k'}[n] \right|^2 = \sum_{k'=0}^{K-1} \left|H^{IRS}_{k,k'}[n] \right|^2 \nonumber\\
             &= \frac{1}{K^2} \sum_{k'=0}^{K-1} \left| H_{l_{IRS}}[n,k'] \right|^2 = \frac{1}{K}\sum_{m=0}^{K-1}\left| h_{l_{IRS}}[n,m] \right|^2\label{eq:ici1},\nonumber
\end{align}
where we used Parseval's theorem for the DFT in the last step.

If the magnitude IRS channel tap is constant, \ie, the malicious IRS modulation results only in phase shifting, \ie, ${| h_{l_{IRS}}[n,m]| =  |h_{l_{IRS}}|}$, this can be simplified further to:
\begin{equation}
    I_{IRS} = \sum_{k' \neq k}\left|H^{IRS}_{k,k'}[n] \right|^2 = |h_{l_{IRS}}|^2= P_{IRS}, \label{eq:ici2}
\end{equation}
which means that the total power received from the IRS, $P_{IRS}$, completely translates into ICI, only. Thus the signal-to-interference ratio~(SIR) due to ICI on the $k^{th}$ subcarrier is given by
\begin{equation}
    SIR_k = \frac{S_k}{I_{IRS}}= \frac{|d_{k}|^2}{|h_{l_{IRS}}|^2 } = \frac{|d_{k}|^2}{P_{IRS}}.
\end{equation}

\end{document}